\documentclass[fleqn,usenatbib]{mnras}

\usepackage{newtxtext,newtxmath}
\usepackage[ampersand]{easylist}
\usepackage{ulem}

\usepackage[T1]{fontenc}
\usepackage{ae,aecompl}

\usepackage{graphicx}	
\usepackage{amsmath}	
\usepackage{amssymb}	

\newcommand{\Msun}{\ensuremath{{\rm M}_{\odot}}}
\newcommand{\g}{$g$}
\newcommand{\z}{$z$}
\newcommand{\Ks}{$K_s$}
\newcommand{\K}{$K_s$}
\newcommand{\gzK}{$gzK_s$}
\newcommand{\PEgzK}{$PE$-$gzK_s$}
\newcommand{\SFgzK}{$SF$-$gzK_s$}
\newcommand{\zs}{$z$$\sim$}

\newcommand{\Mstars}{$M_{\star}$}
\newcommand{\sqdeg}{deg$^2$}

\title[LARgE Survey. I. Stellar Mass Function of Passive Galaxies at $z\sim1.6$]{LARgE Survey. I. Dead Monsters: the Massive End of the Passive Galaxy Stellar Mass Function at  Cosmic Noon}

\author[L.\ Arcila-Osejo et al.]{\hspace{-1.1mm}
Liz Arcila-Osejo$^{1}$\thanks{E-mail: {\tt larcila@ap.smu.ca}},
Marcin Sawicki$^{1}$\thanks{E-mail: {\tt marcin.sawicki@smu.ca}}\thanks{Canada Research Chair},
St\'ephane Arnouts$^{2}$,
Anneya Golob$^{1}$,
\newauthor
Thibaud Moutard$^{1}$,
Robert Sorba$^{1,3}$
\\$^{1}$Saint Mary's University, Department of Astronomy \& Physics and the Institute for Computational Astrophysics, 923 Robie Street, \\ Halifax, Nova Scotia B3H 3C3, Canada
\\$^{2}$Laboratoire d'Astrophysique de Marseille, 38 rue Frederic Joliot Curie, Universit\'e Aix-Marseille, Marseille, F-13388, France
\\$^{3}$Physics Department, Mount Allison University, 62 York Street, Sackville, New Brunswick E4L 1E2, Canada}

\date{Accepted XXX. Received YYY; in original form ZZZ}

\pubyear{2018}

\begin{document}
\label{firstpage}
\pagerange{\pageref{firstpage}--\pageref{lastpage}}
\maketitle
\begin{abstract}
We introduce the largest to date survey of massive quiescent galaxies at redshift \zs1.6. With these data, which cover 27.6~deg$^2$, we can find significant numbers of very rare objects such as ultra-massive quiescent galaxies that populate the extreme massive end of the galaxy mass function, or dense environments that are likely to become present-day massive galaxy clusters. In this paper, the first in a series, we apply our \gzK\ adaptation of the $BzK$ technique to select our \zs1.6 galaxy catalog and then study the quiescent galaxy stellar mass function with good statistics over \Mstars$\sim10^{10.2}$--$10^{11.7}$\Msun --- a factor of 30 in mass --- including 60 ultra-massive \zs1.6 quiescent galaxies with \Mstars$>10^{11.5}$\Msun. We find that the stellar mass function of quiescent galaxies at \zs1.6 is well represented by the Schechter function over this large mass range. This suggests that the mass quenching mechanism observed at lower redshifts must have already been well established by this epoch, and that it is likely due to a single physical mechanism over a wide range of mass. This close adherence to the Schechter shape also suggests that neither merging nor gravitational lensing significantly affect the observed quenched population. Finally, comparing measurements of $M^*$ parameters for quiescent and star-forming populations (ours and from the literature), we find hints of an offset ($M^*_{SF}>M^*_{PE}$), that could suggest that the efficiency of the quenching process evolves with time. 
\end{abstract}






\section{Introduction}\label{sec:introduction}

The rate at which the Universe was forming stars peaked during ``cosmic noon'' at \zs1--3 \citep[see][for a review]{Madau2014}, but some galaxies were already quiescent by that epoch and in some cases perhaps even earlier. In contrast to star-forming galaxies whose numbers increase rapidly with decreasing stellar mass \citep{Sawicki2012hdf, Ilbert2013b, Muzzin2013smf, Davidzon2017}, the most common quiescent galaxies at cosmic noon were those with stellar masses\footnote{We use the \cite{Chabrier2003a} stellar Initial Mass Function (IMF) throughout.} of \Mstars$\sim 10^{10.8}$\Msun\ \citep{Arcila-Osejo2013, Ilbert2013b, Muzzin2013smf, Davidzon2017}. While a second, distinct population of lower-mass (\Mstars$\la 10^{9.5}$\Msun) quiescent galaxies already appears to exist by \zs2 \citep{Tomczak2014}, and could potentially extend to quiescent galaxies with very low masses \citep{Peng2010}, the observed low-mass quiescent galaxies at these redshifts are few in number compared to their more numerous massive (\Mstars$\sim 10^{10.8}$\Msun) cousins that dominate the quiescent population both by number and by integrated stellar mass.  

A significant number of such massive (\Mstars$\sim 10^{10.8}$\Msun) quiescent galaxies is now spectroscopically confirmed at $1\la z\la 3$ \citep{Dunlop1996, Cimatti2004, Glazebrook2004, Cimatti2008, McCarthy2008, Gobat2012, Onodera2012, Belli2014z1, Belli2014z2,  Belli2016} and their existence provides an important test of galaxy evolution models \citep[e.g.,][]{Somerville2012, Merson2013, Wellons2015, Behroozi2018}. However, a handful of spectroscopically-confirmed examples of even more massive (\Mstars$\sim 10^{11.5}$\Msun) quiescent galaxies is now known \citep[e.g.,][]{Onodera2012, Belli2016, Kado-Fong2017} and given their extreme masses and very early inferred formation and quenching times ($z\ga4$, \citealt{Belli2018}; see also \citealt{Pacifici2016, Glazebrook2017}) they could provide even more stringent tests of our galaxy formation models.  Moreover, ultra-massive passiveley evolving galaxies (UMPEGs) at cosmic noon, which for concreteness we henceforth define to be at $1<z<3$ and to have \Mstars$>10^{11.5}$\Msun,  could be related to other interesting galaxy populations, including massive dusty star-forming galaxies at higher redshifts and cluster brightest central galaxies (BCGs) in clusters at lower redshifts. 

UMPEGs could have of course assembled from previously quenched lower-mass progenitors such as are found in some \zs2 clusters \citep{Gobat2013, Newman2014}, or formed via the merger-induced quenching of merging star-forming galaxies found in, say, high-$z$ protoclusters  \citep[e.g.,][]{Steidel2000, Capak2011, Cucciati2014,  Oteo2018}. However, if UMPEGs formed directly through the quenching of star formation in equally  {\it{ultra-massive star-forming}} galaxies, then these star-forming \Mstars$>10^{11.5}$\Msun\ progenitors would have to have had star formation rates (SFRs) of 200-1000 \Msun/yr if, just prior to their quenching, they were on \citep{Whitaker2012, Whitaker2014} or above \citep{Elbaz2018} the $z\ga2$ star-forming main sequence \citep[SFMS, ][]{Daddi2007,Elbaz2007,Noeske2007,Sawicki2007}. Such high SFRs and stellar masses are typical of Submillimetre Galaxies (SMGs) at $z\ga2$ \citep[e.g.,][]{Michalowski2012, Michalowski2014}, or super-Main Sequence massive starbursts \citep{Elbaz2018}, suggesting that massive dusty star-forming galaxies could be the direct progenitors of high-$z$ UMPEGs. 

The stellar masses of UMPEGs are also consistent with the stellar masses of the brightest cluster galaxies (BCGs) of massive clusters at low redshift \citep[\Mstars $\sim 10^{11.7}$ \Msun, with a few examples at \Mstars$>10^{12}$ --- see compilation by ][]{Lidman2012}. Central cluster galaxies of similarly high masses are also known to \zs1 and, for a few cases, to \zs1.5 \citep{Stott2010, Lidman2012}. Additionally, BCG formation models suggest that cluster massive central galaxies may have already been quiescent and very massive (\Mstars$\sim$ a few $\times 10^{11}$) by \zs 1.5 \citep{DeLucia2007}. Altogether, these comparisons suggest that UMPEGs could be the progenitors of present-day Brightest Cluster Galaxies (BCGs). 

To further test these scenarios we need to compare the number densities and clustering strengths of UMPEGs with those of the other galaxy populations of interest. Additionally, the detailed shape of the stellar mass function (SMF) of quiescent galaxies can carry in it information about the quenching process that shuts down star formation \citep[e.g., ][]{Peng2010, Peng2012} making the shape of the quiescent galaxy SMF a test of quenching models. However, at present even the number density of the most common quiescent galaxies at \zs2, those with \Mstars$\sim 10^{10.8}$\Msun, is not yet satisfactorily constrained by observations given both the cosmic variance affecting the relatively small, degree-scale fields used in $z>1$ mass function studies as well as differences in how samples are selected (see \S~\ref{sec:SMF}). The number density of the {\it{truly}} massive UMPEGs is even more poorly constrained because the quiescent galaxy stellar mass function appears to drop rapidly at very high masses.  Our goal then is to measure the SMF of high-$z$ quiescent galaxies over a large mass range, including very high UMPEG masses.

In constraining the number density of UMPEGs we face two problems: (1) UMPEGs are rare,  while (2) the selection of high-$z$ quiescent galaxies requires deep near-infra-red imaging and very deep optical imaging.  Consequently, to date, at high redshift ($z\ga1.5$) studies have been limited to a handful of degree-scale --- or even smaller --- fields \citep{Hartley2008, McCracken2010, Furusawa2011, Arcila-Osejo2013, Ilbert2013b,Muzzin2013smf, Ishikawa2015}, some of which, such as the COSMOS field, are duplicated between studies. These studies do suggest, however, that the number density of \Mstars $> 10^{11.5}$\Msun\ UMPEGs at \zs1.5--2.5 is $\la 10^{-5}$Mpc$^{-3}$, albeit with large statistical uncertainties, significant but poorly constrained cosmic variance given the very biased regions that the UMPEGs can be expected to reside in, and systematic differences due to different sample selection methods. These low UMPEG number densities, one twentieth or less those of the most common $z\sim2$ passive galaxies around \Mstars$\sim 10^{10.8}$\Msun, indicate that to study this rare population we need to survey areas much larger than the degree-scale surveys carried out to date. 

To better understand both  the evolutionary pathway of UMPEGs and their relationship to other populations at high and low redshift, we need to first identify a statistically significant sample that is embedded in a full range of environments and is free of cosmic variance.  This calls for a dataset that covers an order of magnitude more area than previous studies. With this sample, it will be possible to constrain the poorly-known number density of high-$z$ UMPEGs, to relate their stellar masses to those of their dark matter halos through clustering analysis, and to examine the environments in which they reside. These are the goals of our program, which we call the Large Area Red $gzK$ Environments (LARgE) Survey, and in which we set out to find and then study the properties of a large sample of \zs1.6 ultra-massive passive galaxies. In this paper, the first in the LARgE series, we describe the dataset we use and the resultant sample of colour-colour-selected high-$z$ galaxies, and present their number counts and passive galaxy stellar mass function. In subsequent papers in the series we will measure the masses of the dark matter halos of the most extreme quiescent galaxies via galaxy-galaxy clustering (Gurpreet Kaur Cheema et al., submitted to MNRAS), constrain the growth of these galaxies via galaxy-galaxy mergers (Liz Arcila-Osejo, in prep.), and identify a sample of massive galaxy cluster candidates at \zs1.6 identified by the presence of massive quiescent galaxies (Liz Arcila-Osejo et al., in prep.).

Throughout this work, and unless otherwise noted, we use the AB magnitude system \citep{Oke1974}, the ($\Omega_M$, $\Omega_\Lambda$, $h$) = (0.3, 0.7, 0.7) cosmology, and the \cite{Chabrier2003a} stellar initial mass function (IMF).

\section{Data and Catalogs}\label{sec:data}
 
Our project makes use of archival optical and NIR imaging, primarily from the Canada France Hawaii Telescope (CFHT):   The Deep and Wide fields of the CFHT Legacy Survey (CFHTLS) for the optical data, and CFHT WIRCam IR imaging from the WIRDS and VIPERS-MLS projects for the IR data; a small amount of IR imaging in the WIRDS dataset comes from UKIRT.  We perform object detection in the $K_s$-band images over these fields, do matched-aperture photometry in other bands, and finally select high-$z$ galaxies using an adaptation of the $BzK$ technique to our filter set. 
 
In the following sections we first describe these datasets and then detail  how we used them to construct our $K$-band selected catalogs and select high-$z$ galaxies.

\subsection{Data}

 \begin{figure*}
	\includegraphics[width=1.0\textwidth]{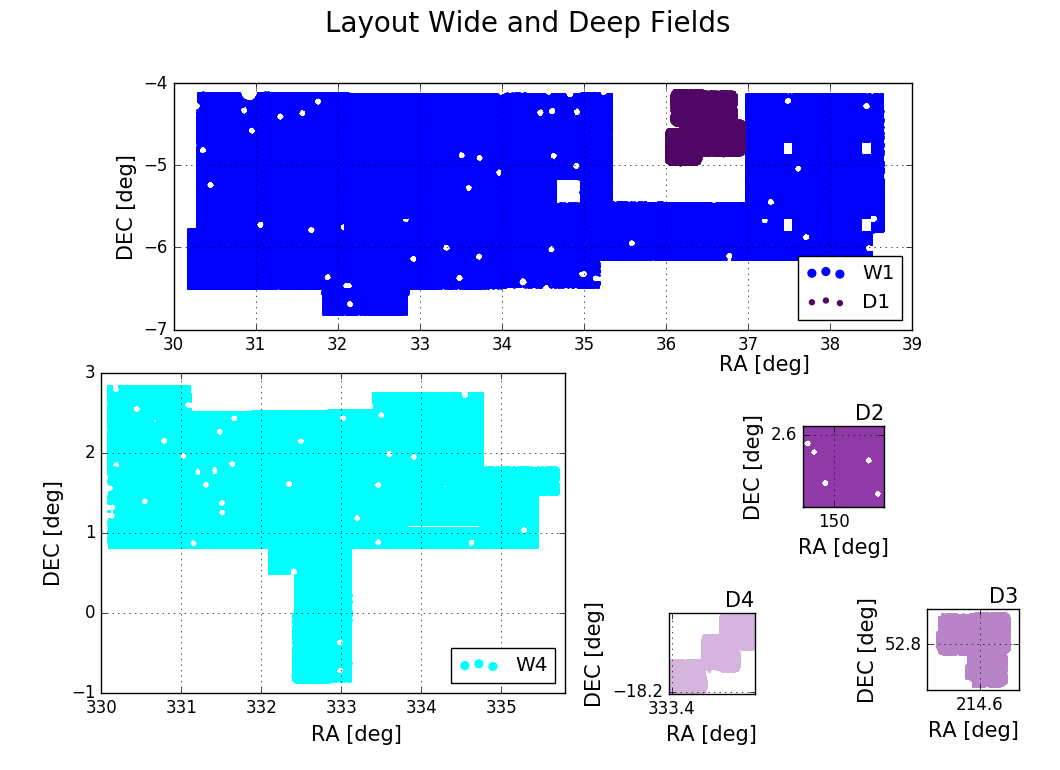}
    \caption{Layout of our fields.  The Wide fields are shown in shades of blue and the Deep fields in purple.  Note that the D1 field is inside the W1 panel.  Gaps represent either areas where there are no K\protect\textsubscript{s} observations
available or where the area has been masked due to the presence of
bright stars or artifacts.}
    \label{fig:Layout}
\end{figure*}
\subsubsection{Optical Data: The CFHT Legacy Survey\label{sub:Optical-Data}}

The Canada-France-Hawaii Telescope Legacy Survey (CFHTLS) is a large public project composed of two main surveys: Deep and Wide. The CFHTLS Deep component consists of four independent, very deep, one square degree pointings labelled D1, D2, D3 and D4. The CFHTLS Deep surve

y was carried out to detect type Ia supernovae and study the galaxy population to faint magnitudes.  CFHTLS-Wide is divided into four fields (W1--W4) and its purpose  is to study matter distribution, large scale structure and clusters of galaxies. For our work, we used all four Deep fields (D1--D4) and two of the Wide fields (W1 and W4) of the CFHTLS. 

CFHTLS  was carried out using MegaCam, the wide-field optical imager at MegaPrime. Each MegaCam CCD pointing covers a  $\sim$1~$\deg$ $\times$ $\sim$1~$\deg$  field of view with a sampling of 0.186 arc seconds per pixel. All CFHTLS fields were observed in five broad-band filters ($u{*}, g', r', i'$, (or its replacement, $i2'$), and $z'$).  From brevity we will drop the superscripts and use $ugriz$ to designate this MegaCam filter set. In our work we primarily make use of the \g\ and \z\ filters, as described in Section~\ref{sec:gzkSelection}.  For the Deep fields, the 80$\%$ completeness limit for extended objects in g-band is 25.29, 25.30, 25.29 and 25.26 and for the z-band it is 23.83, 23.90, 23.71, 23.76 AB mags respectively for D1, D2, D3 and D4 (\citealt{Cuillandre2010}). Likewise, for the Wide fields, W1 and W4, in the g band the completeness limit is 24.67 and 24.71, and in the \z\ band is 22.91 and 22.90, respectively (\citealt{Hudelot2012a}).

CFHTLS data processing was done at Terapix\footnote{http://terapix.iap.fr/}. Pre-processing involved image quality checking, flat fielding, stacking from dithering, identification of bad pixels, removal of cosmic rays and saturated pixels, background estimation, and astrometric and photometric calibration. Finally, Terapix is also involved in archiving and providing data products to the scientific community.  For the Deep fields we used the Terapix T0006 processed and calibrated stacks, while for the Wide fields, we use the T0007 release.  One of the key differences between these two releases is a different photometric calibration and, be comparing data in the T0006 and T0007 versions of the Deep fields we noticed that there was a systematic calibration offset in the z-band, such that $z_{T0007} = z_{T0006} + 0.1$ \citep[see also ][who reported a shift of 0.148 $\pm$0.054]{Moutard2016a}.  We thus applied a 0.1 mag offset to the \z-band  zeropoints in the Wide data (for which we used the T0007 release) in order to make these data consistent with the Deep fields (for which we used the T0006 dataset).  We note here that we found no systematic shift in the other optical bands.

\begin{table*}
\label{tab:numberCountsAllObjects}.
\caption[Summary of detected sources in all Deep and Wide fields.]{Summary of detected sources in all Deep and Wide fields. The area represents the effective area once external flags have been taken into account. The total number of sources includes all galaxies, irrespective of redshift and stars to a limiting magnitude of 23.5 for the Deep Fields and 20.5 for the Wide Fields. \gzK\ sources are high-redshift objects (z$\sim$1.6) that have been identified either as star-forming or passive based on their location on a (g-z) vs. (z-K$_{s}$) colour-colour plot (see Section \ref{sec:gzkSelection}). The Depth in this table represents the completeness limit of our sample given the nature of our sources (star-forming or passive). If a range is given, the lower value indicates \gzK\ galaxies that are unambiguously selected as star-forming galaxies at $z\sim1.6$, while the higher value includes possible low-redshift interlopers. }
\begin{centering}
\begin{tabular}{|c|r|c|c|c|c|}
\hline 
Field & area  & No.\ all objects & No.\ \gzK\ galaxies & depth SF-\gzK & depth PE-\gzK\\
 & {[}$deg^{2}${]} & & & \Ks(AB) & \Ks(AB)\\
\hline 
D1 & 0.69 & 71712 & 16517--18804 & 23.5 & 23.5\\
D2 & 0.91 & 109490 & 16821--22897 & 23.5 & 23.0\\
D3 & 0.45 & 48425 & 10490--11795 & 23.5 & 23.5\\
D4 & 0.46 & 48757 & 10160--12090 & 23.5 & 23.5\\
W1 & 15.53 & 236213 & 5502 & 20.5 & 20.5\\
W4 & 9.56 & 191570 & 3254 & 20.5 & 20.5\\
\hline 
\end{tabular}
\par\end{centering}
\end{table*}

\subsubsection{Infrared Data: WIRDS and VIPERS-MLS\label{sub:Infrared-Data}}

To complement the Deep survey optical data we used the WIRCam Deep Survey \citep[WIRDS;][]{Bielby2012}. WIRDS was a large project carried out during 2006 to 2008 to obtain near infrared, broadband imaging for the four CFHTLS Deep fields. The results for this survey included images from the J, H and K\textsubscript{s} filters that have been re-sampled to match the pixel scale of MegaCam. 
Most of the WIRDS  were taken with the Wide-field Infrared Camera imager (WIRCam) at the CFHT, although the $J$-band imaging of the D2 field was obtained using the Wide Field Camera (WFCAM) instrument on the United Kingdom Infrared Telescope (UKIRT).  Most of the effective areas for the Deep fields are less than one square degree (with the exception of D2). We used the T0002 WIRDS data release which we obtained from Terapix. The WIRDS  50\% completeness limits for point-like objects is reached at $K$=24.5 AB,  except for the D2 (COSMOS) field which reaches $K$=24.0 AB  (\citealt{Bielby2012}) .

For the Wide fields we used the \Ks-band images taken and processed as part of the Visible Multi-Object Spectrograph (VIMOS) Public Extragalactic Multi-Lambda Survey or VIPERS-MLS\footnote{http://cesam.lam.fr/vipers-mls/}  (\citealt{Moutard2016a}). These data were also obtained with WIRCam on CFHTLS and were  processed into square-degree patches that match in size and pixel scale the optical pointings of the CFHTLS-Wide.  These data were designed to reach  a $80\%$ completeness limit of K\textsubscript{s }$\leq$ 22.0 AB. The K\textsubscript{s}-band calibration of VIPERS-MLS matches that of the WIRDS Survey and also matches the VIDEO and UKIDSS calibrations where overlap exists \citep{Moutard2016a}.

\subsection{K-selected Catalogs}\label{sec:Ks-selected-Catalogs}

\subsubsection{Effective areas}

In Figure \ref{fig:Layout} we show the layout of our g, z and \Ks\ coverage in W1 and W2 and \g, \z, $H$ and \Ks\ coverage in D1 to D4. We masked areas where object detection and/or photometry could be affected by the presence of bright star halos or other artefacts. For the Wide optical images, external flags of bright stars and cosmic ray trails were taken from \citealt{Erben2012}; we performed visual inspection of each one of these masks to ensure the masking was correct. For the corresponding Wide-field K\textsubscript{s} band images we created our own external masks by visual inspection. Similarly, for the Deep Fields, both optical and infrared bands, we constructed our own external flags by visual inspection of the different images. After masking, our total remaining area is 27.60~deg$^2$, including 25.09~deg$^2$ in the Wide fields and 2.51~deg$^2$ in the Deep fields. The areas of the individual fields, after masking of bright stars and other unusable subareas, are given in Table~1.

\subsubsection{Photometry}

Because quiescent galaxies at high redshift are expected to be very red ($g-K_s \gtrsim 5$~AB), it is critical to detect sources in the reddest available band ($K_s$ in our case) as optical detection could miss a large number of such objects.  With this in mind we used the SExtractor package (\citealt{Bertin1996}, version 2.19.5) to detect objects in the \Ks-band images and then perform forced-aperture photometry in the other bands. 

A source was identified in the K\textsubscript{s} image based on it having at least five contiguous pixels above a detection threshold.  The Deep-fields images are much deeper than the Wide and suffer from more correlated noise, which means that threshold may not be the same in Deep and Wide. By trial and error and visual inspection of SExtractor detections, we determined and subsequently used a threshold requirement of five contiguous pixels each above 1.5$\sigma_{sky}$ in the Wide areas, and 1.2$\sigma_{sky}$ in the Deep. Photometry is then performed at the \Ks-band positions in \g, \z, and $H$ (in the Deep fields), using SExtractor's dual-image mode, as explained below. 

In all the fields, total \Ks-band fluxes were calculated using auto mags (total magnitudes from a Kron-like apertures). The Deep and Wide fields were then treated differently when computing colours. In the Deep fields, which we used to push into the faint limit of the galaxy population, we wanted to maximise depth. For this reason we determined colours in fairly small (10 pix diameter, or 1.86'') apertures on \g, \z, and $H$ images whose PSF had been homogenised to match the \Ks-band PSF (\citealt{Sato2014}). This approach follows exactly, and on the same data, that given in \cite{Arcila-Osejo2013}; it gives high S/N in the colours while ensuring the fluxes are measured over the same physical areas of the objects despite the small apertures.  In the Wide fields, which we use to study the bright end of the population, we did not make PSF-homogenised images, but, instead, we used SExtractor's total magnitudes (measured over large apertures determined individually for each object in the \Ks-band) to measure colours; because for our relatively bright objects these total magnitude apertures are significantly larger than the PSF, this approach gives accurate colours without unacceptably degrading the S/N. 

Finally, the photometry of each object was corrected for foreground Galactic dust.  This was done using the 
\cite{Schlegel1998} Galactic dust maps,  which are based on COBE and IRAS 100 $\mu m$ to 240 $\mu m$ emission maps, but using the updated reddening curves of \cite{Schlafly2011} that are based on stellar spectra in the Sloan Digital Sky Survey (SDSS).

The result is a catalogue that includes $\sim$700,000 \Ks-band selected objects in an effective area (i.e., after masking) of 27.60~deg$^2$. See Table~1 for the breakdown by field.


\section{High-z Object Selection and Number Counts}\label{sec:numberCounts}

\subsection{Selection of gzK galaxies}\label{sec:gzkSelection}

\citet{Daddi2004b} developed a simple technique that uses the $B-z$ vs.\ $z-K$ colour-colour plane for selecting high-redshift galaxies and classifying them as either Passively Evolving ($pBzK$) or star-forming ($sBzK$). The technique is grounded in noting how galaxy spectral models (\citealp{Daddi2004b} used \citealp{Bruzual2003} models attenuated with \citealp{Calzetti2000a} dust) place in the $BzK$ colour plane. Spectroscopy, along with morphological studies, support the validity of $BzK$-selected passive and star-forming galaxies (\citealt{Daddi2004b, Daddi2005, Ravindranath2007, Hayashi2009, Onodera2010, Mancini2010a}), and the simplicity and relative frugality of the technique (only three passbands are needed) continues to make it a popular choice for selecting star-forming and quiescent samples of distant galaxies \citep[e.g.,][]{McCracken2010, Furusawa2011, Yuma2011, Kurczynski2012, McCracken2012, Yuma2012, Lee2013, Rangel2013, Sommariva2014, Fang2015, Ishikawa2015, Ishikawa2016}.

 \begin{figure}
	\includegraphics[width=\columnwidth]{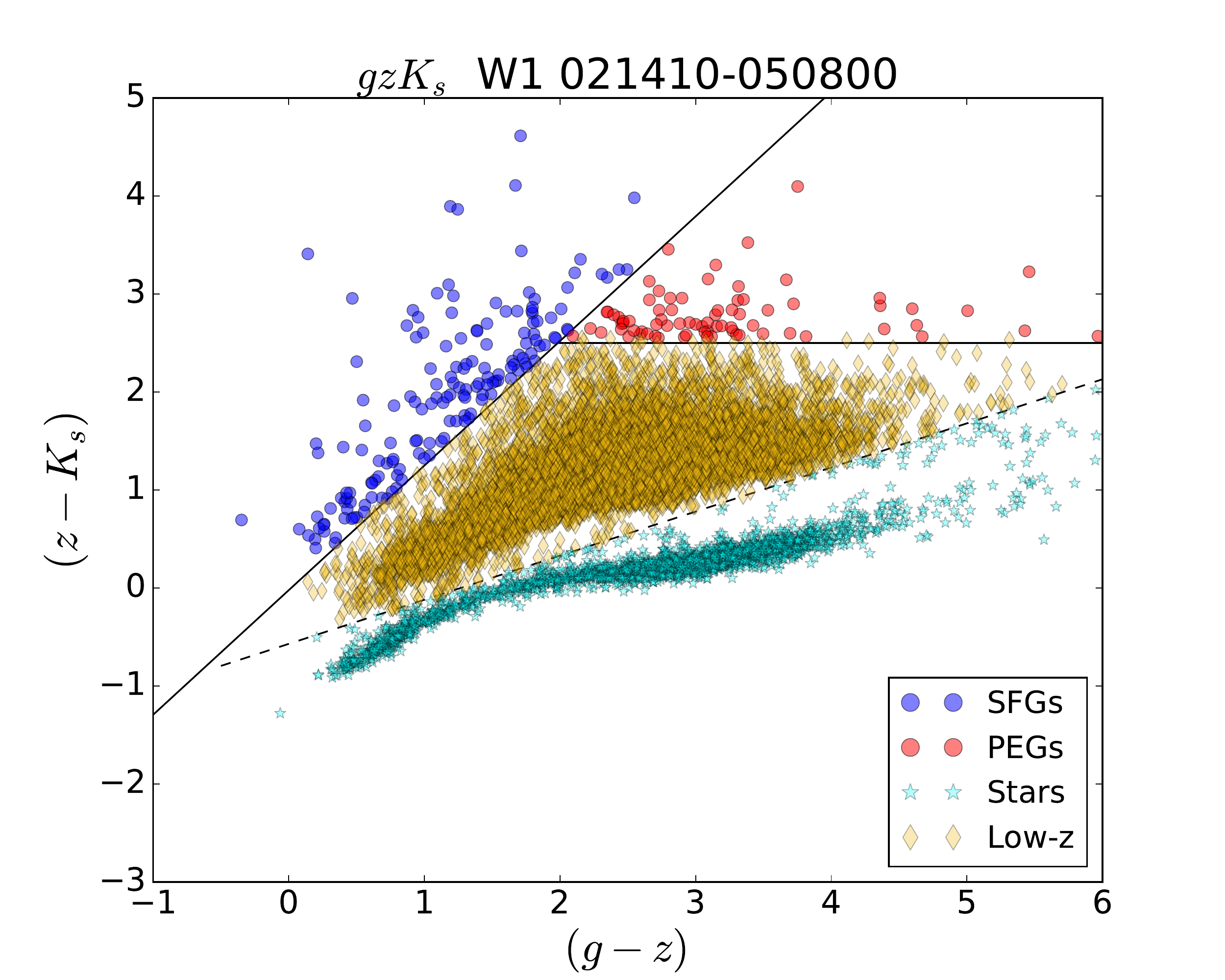}
    \caption{Selection of star-forming and passive galaxies down to $K_s < 20.5$ in the $(g-z)$ vs.\ $(z-K_s)$ plot for one $\sim$1~deg$^2$ patch in the Wide fields.  This represents $\sim$3\% of our Wide dataset.  Blue symbols represent $gzK$ galaxies that were classified as star-forming, red symbols are for passive galaxies, orange for low-redshift galaxies and cyan for stars, classified as described in the text.}
    \label{fig:gzKdiagram}
\end{figure}

In \citet{Arcila-Osejo2013} we adapted the classic $BzK$ technique to the \gzK\ bandbasses available in CFHTLS fields by noting how the \citet{Daddi2004b} spectral synthesis models shift when redrawn in the  \g$-$\z\ vs. \z$-$\Ks\ plane (the shifts are small given the similarity of the filter sets). For completeness, we give here the equations from \citet{Arcila-Osejo2013} for selecting high-$z$ galaxies using this approach, noting again that --- for brevity --- hereafter we use $g$ and $z$ to denote the CFHT Megacam $g'$ and $z'$ filters. These $gzK_s$ selection criteria are
\begin{equation}
( z-K_{s} ) - 1.27 (g-z) \geq - 0.022,
\label{eq:sf-selection}
\end{equation}
for star-forming galaies, and
\begin{equation}
(z-K_{s})-1.27(g-z)<-0.022\;\ \cap\:\:(z-K_{s})\geq2.55,
\label{eq:pe_select}
\end{equation}
for passive galaxies \citep{Arcila-Osejo2013}.  

To select and classify high-redshift galaxies we then apply the cuts described by equations~\ref{eq:sf-selection} and \ref{eq:pe_select} to our Wide catalogues. This is illustrated in Figure~\ref{fig:gzKdiagram} for a $\sim$ 1 deg$^2$ subarea of our Wide catalogue (i.e., approximately 3\% of our dataset).  In common with the classic $BzK$ selection, the upper-left region of the diagram contains star-forming high-$z$ galaxies (\SFgzK, blue points) and the upper right has quiescent high-$z$ galaxies (\PEgzK, red points).  Stars follow a distinct sequence in the lower part of the diagram (green points).  Yellow points mark lower-$z$ galaxies. 

In the Deep fields the \gzK\ technique can identify high-redshift galaxies as well, but by itself it is not sufficient to distinguish between star-forming and passive sub-classes. This is because to separate \PEgzK\ from \SFgzK\ galaxies we need to be able to measure colours to (\g $-$ \Ks)$>$4.5 and in the Deep fields the \Ks-band data are so deep that some \gzK\ galaxies are undetected in the \g-band. While clearly at high redshift (because they have red \z-\Ks\ colours), we cannot tell their evolutionary state from \gzK\ data alone.  Instead, following \citet{Arcila-Osejo2013}, we use these galaxies' $H-K_s$ colours to break the degeneracy.  While in the absence of a \g-band detection it is not clear whether the red \z-\Ks\ colour is due to dust or the presence of the Balmer/4000\AA\ break complex,the $H-K_s$ colour can remove the age-dust degeneracy, as shown in \citet[see their Fig.~3]{Arcila-Osejo2013}. For completeness, we give here the relation from \citet[their Eq.~6]{Arcila-Osejo2013}, 
\begin{equation}
(z-H) > 2.4(H-K_s)+1,
\end{equation}
which identifies quiescent galaxies from amongst \g-undetected \gzK\ candidates. 

Applying \gzK\ selection to the Wide fields and the \g\z\Ks+$H$ method to the Deep, we then construct catalogues of star-forming (\SFgzK) and quiescent (\PEgzK) galaxies. We limit our catalogue depth in the \Ks-band in each of the fields, as given in Table~2 in order to avoid incompleteness and -- in the Wide fields -- to ensure that the vast majority of objects are detected in the \g-band. 

To check that our \gzK-selected objects are not point sources but are resolved --- and therefore likely to be galaxies --- we compared their peak \K-band surface brightnesses (PSBs) with total magnitudes.  Stars are compact (high PSB for their total magnitude) as compared to galaxies and present a clear and distinct locus in a PSB-mag diagram. All of our \PEgzK\ objects with K$_{s}<18.5$ (a total of five sources) are consistent in the PSB-vs-mag diagram with point source morphologies and we excluded them from further consideration. In comparison, only $\sim7\%$ of \PEgzK\ objects detected at $18.5\leq K_{s}<19.5$ and less than $0.5\%$ of \PEgzK\ objects detected at $19.5\leq K_{s}<20.5$ are point sources in our data; they, too, have been removed from further analysis. 

We also visually inspected the images of all \PEgzK\ objects with $18.5 < K_s < 20$ in the Deep and Wide fields ($\sim$600 objects). Only six of these $\sim$600 objects (i.e., $\sim$ 1\% of the total) were found to be corrupted (edge effects, closeness to bright-star diffraction spikes, corrupted pixels, etc.) and were removed from the sample. 

A closer inspection of ultra-bright galaxies ($K_{s}<19.5$; 73 objects) revealed the presence of six possible late-stage major mergers: Galaxies with double cores that were not properly segmented by SExtractor.  By visually identifying a boundary between the two sources, we recovered a separate background-corrected flux for each source in the pair, confirming the fact that what appears to be an ultra-bright passive galaxy could in fact be an ongoing, major merger between two fainter ($K_{s}\sim20$) sources. We retained these six objects in our catalogue. These major merger candidates could potentially result, with time, in an ultra-bright passive galaxies. 

Finally, we matched all K$_{s}<20$ objects with the
Chandra\footnote{http://cda.harvard.edu/chaser/}, XMM\footnote{http://nxsa.esac.esa.int/nxsa-web/},
NED\footnote{https://ned.ipac.caltech.edu/forms/nnd.html} and Simbad\footnote{http://simbad.u-strasbg.fr/simbad/sim-fbasic} databases in order to discard any possible AGN sources in our sample. In the Deep fields, no X-ray sources were found in the the XMM and Chandra databases but one object in D3 was identified in the Simbad catalogue as an AGN source in the Extended Groth Strip, belonging to the AEGIS-X Deep Survey \citep{Goulding2012}.  This object was removed from our sample, as were two X-ray sources in field W1. 

At the end of this process we were left with a large sample of \PEgzK\ and \SFgzK\ objects, all likely to be high-redshift galaxies.  The numbers are listed in Table~2.

\subsection{Number Counts}\label{NumberCounts}

To study the number counts of \gzK\ galaxies we used the Wide fields for the bright end and the Deep fields for the faint end.  We used simulations \citep{Arcila-Osejo2013,Sato2014} to determine incompleteness corrections in the Deep fields, while in the Wide fields we truncated our analysis below magnitudes where the Wide field number counts diverge from the Deep fields data.

 \begin{figure}
	\includegraphics[width=\columnwidth]{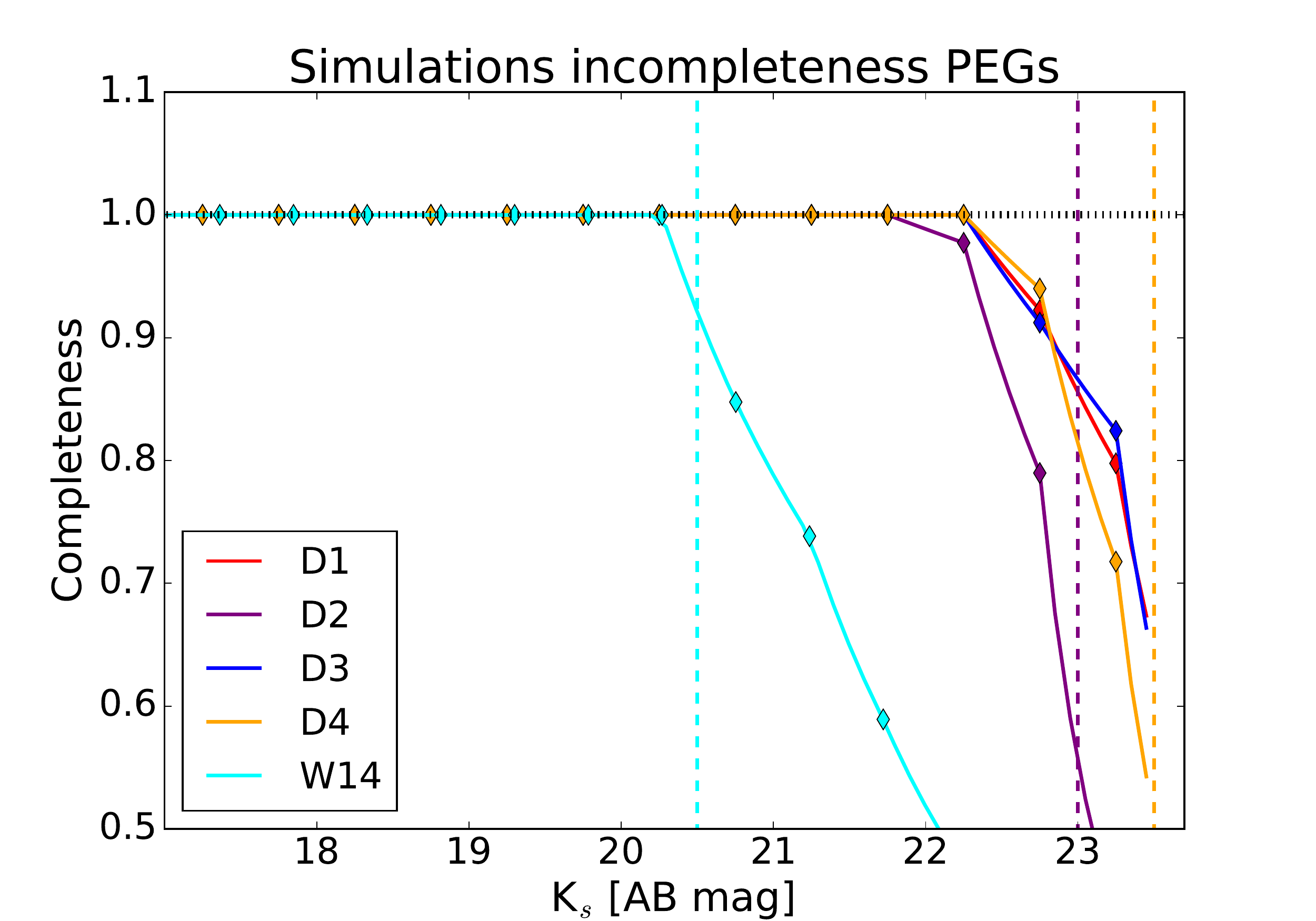}
    \caption{Detection completeness for \PEgzK\ galaxies.  For the Deep fields the completeness was determined using recovery rates of artificial objects implanted into the data.  For the Wide fields completeness is defined as the ratio of the Wide-field number counts to the Deep-field number counts.  The vertical lines indicate our adopted survey limits for the various fields: W1 and W4 (cyan), D2 (purple) and D1, D3, and D4 (orange).}
    \label{fig:completeness-vs-K}
\end{figure}

 \begin{figure}
	\includegraphics[width=\columnwidth]{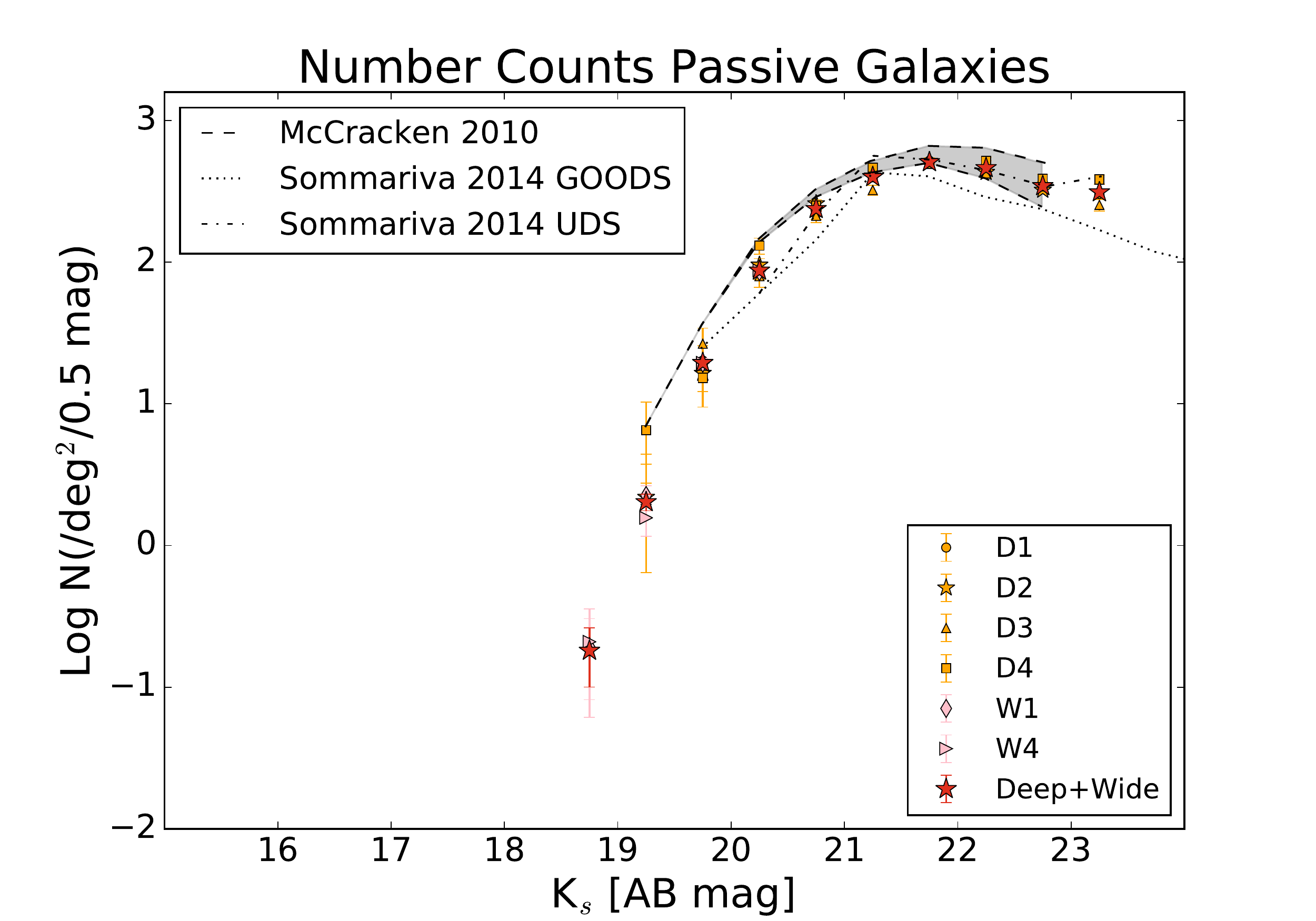}
    	\includegraphics[width=\columnwidth]{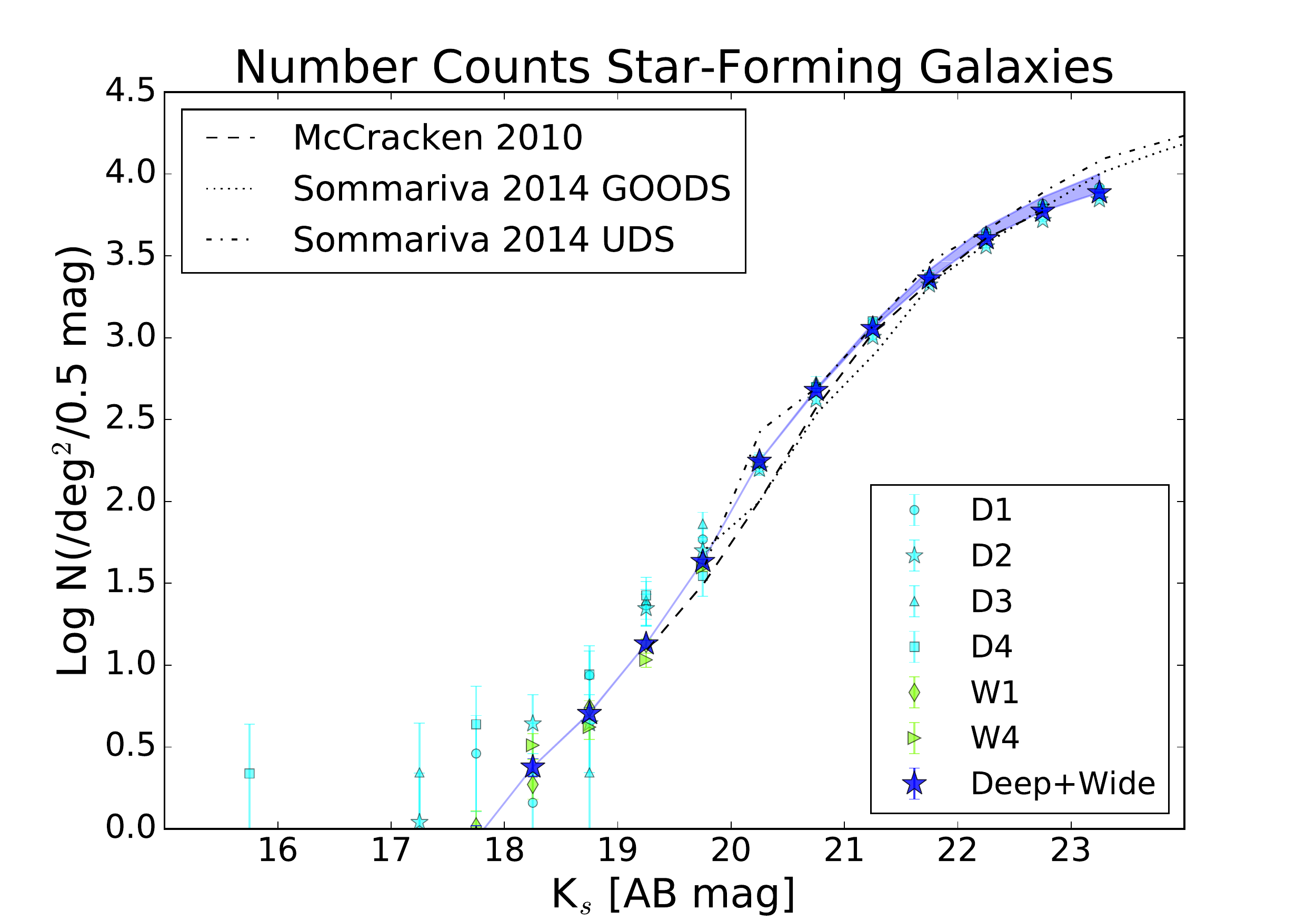}
        \caption{Number counts of passive (top) and star-forming (bottom) \gzK\ galaxies. Individual fields (D1--D4, W1, W4) are shown with small symbols. Large star symbols show the combined number counts. The shaded blue area in the lower (SF-\gzK) panel represents the upper/lower limits obtained when including/excluding \g-undetected objects in the Deep fields. Deep-field number counts have been corrected for incompleteness using simulations, while for the Wide fields incompleteness is negligible to the depth plotted. Our results are in good agreement with other authors although, given our larger survey area, we have better statistics and go to significantly brighter magnitudes. 
}
    \label{fig:numberCounts}
\end{figure}

The incompleteness simulations in the Deep fields are described in detail in \citet{Sato2014}. Briefly, artificial objects were added at random locations into the science images with empirically motivated morphological parameters. High-z star-forming galaxies were assumed to be disk-like objects with effective radii between 1 $\leq$ $r_{e}$ $\leq$ 3 kpc \citep{Yuma2011} while passive galaxies present a more compact morphology and effective radii between 3 $\leq$ $r_{e}$ $\leq$ 6 kpc \citep{Mancini2010a}. Once artificial galaxies are added into an image, SExtractor and catalogue-making are run with the same settings as were used for creating the object catalogue (Sec.~\ref{sec:data}) and the recovered/input numbers of simulated galaxies give detection completeness as a function of magnitude. The results are shown for \PEgzK\ galaxies in 
Fig.~\ref{fig:completeness-vs-K} (a similar result was obtained for SF-\gzK\ galaxies but is not shown).  Incompleteness corrections are simply 1/completeness and are applied to the observed number counts in the Deep fields.  The resulting, incompleteness-corrected number counts are shown in Fig.~\ref{fig:numberCounts}. 

Incompleteness corrections for the Deep fields are typically  \ensuremath{\sim}1.02 over K$_{s}$ = 17 \textminus{} 22 (AB) mag but become larger at fainter levels. We limited our analysis in magnitude bins where incompleteness corrections are smaller than a factor of two, namely $K_{s}$ $<$ 23.5 mag (except for D2, where our passive population is only complete to $K_{s}<23.0$). For the Wide fields, instead of carrying out simulations we simply stop the analysis at magnitudes where the observed Deep-fields number counts diverge from the Wide-fields number counts.  This happens at $K_s = 20.5$, shown in Fig.~\ref{fig:completeness-vs-K} with the vertical dashed line.  This approach allows us to constrain the bright end of the number counts without needing to go into computationally expensive simulations.


 \begin{figure}
    	\includegraphics[width=\columnwidth]{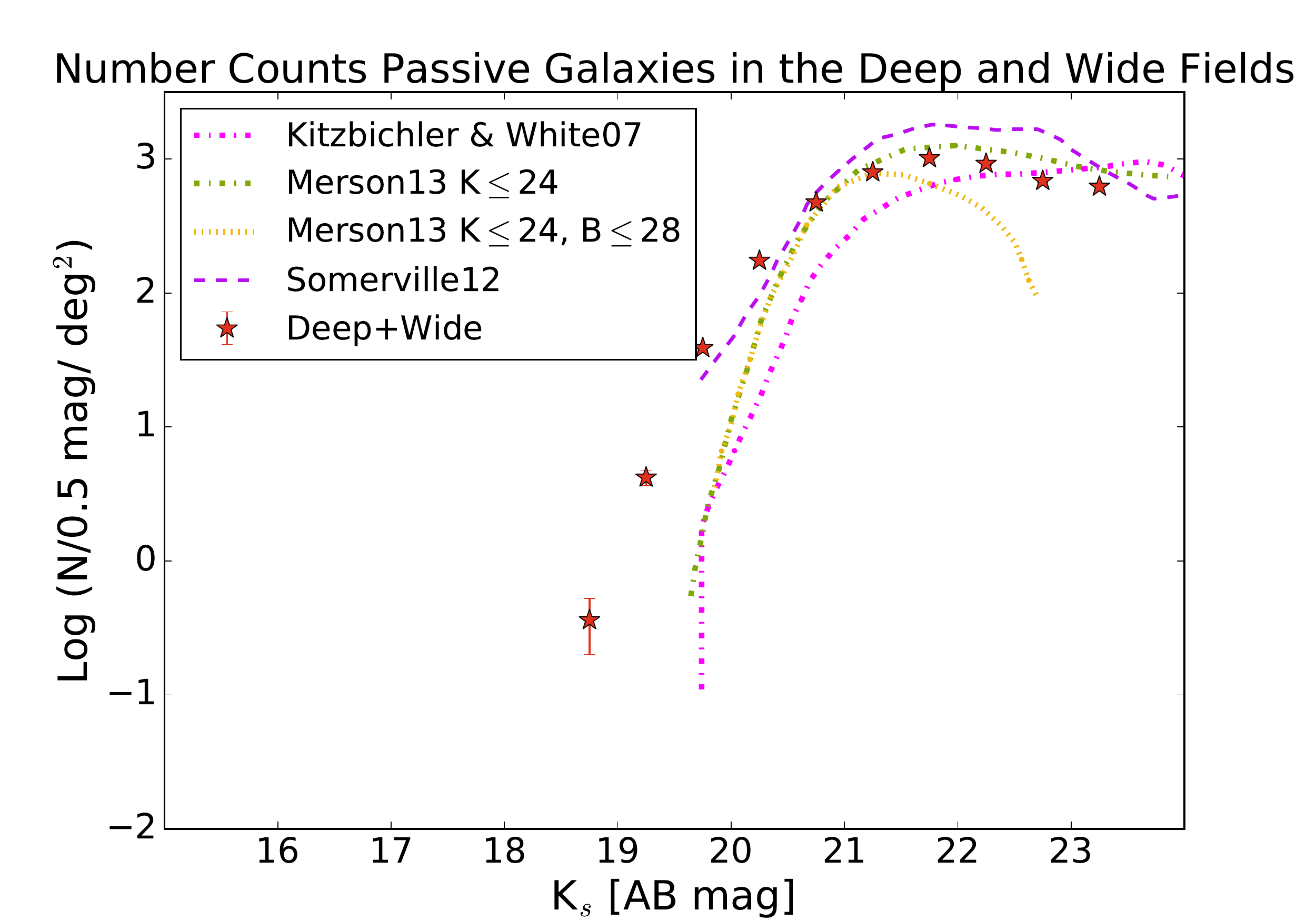}
    \caption{Comparison of our observed PEG number counts with model predictions. 
}
    \label{fig:numberCountsVsModels}
\end{figure}

The number counts of \PEgzK\ and \SFgzK\ galaxies (corrected for incompleteness at the faint end in the Deep data) are shown in Fig.~\ref{fig:numberCounts}. We show each independent field (D1--D4, W1, W4) with a different symbol.  Error bars are Gaussian $\sqrt{N}$ uncertainties. The small variations between our fields are most likely due to cosmic variance. Our total number counts are also presented in Table~2.

For both star-forming and passive galaxies, our results are in good agreement with the recent $BzK$ number counts of \cite{McCracken2010} and \cite{Sommariva2014}, as shown in Fig.~\ref{fig:numberCountsVsModels}. However, we have better statistics and go significantly brighter in the population.

\subsubsection{Comparison with models}

Figure \ref{fig:numberCountsVsModels} shows a comparison between the observed and predicted number counts from semi-analytical models of passive $BzK$ galaxies. Shown in red points are our observations and for comparison, we plot three semi-analytical models: \citet{Kitzbichler2007}, \citet{Merson2013}, and \citet{Somerville2012}.

\cite{Kitzbichler2007} compared observations of the high-redshift population with predictions from the Millennium simulation (\citealt{Springel2005}), which follows the hierarchical growth of dark matter structures from z=127 to the present, including several physical processes such as star-formation, gas cooling, growth of super-massive black holes, stellar population synthesis modelling for photometry and a new treatment for radio mode feedback only for galaxies at the centre of groups or clusters (hence not applied uniformly to all massive galaxies). \cite{Somerville2012} semi-analytical models, similar to those \cite{Kitzbichler2007}, include hierarchical dark matter growth, star-formation, gas cooling,  supernova and AGN feedback, and metal enrichment. Feedback processes are mainly driven by supernovae at the faint end of the SMF and by AGN feedback at the massive end.  Finally, \citet{Merson2013} populate dark matter halos from the Millennium simulation with galaxies using the GALFORM semi-analytical prescription (\citealt{Bower2006}). This prescription includes gas cooling, supernova and AGN feedback (only effective in systems with quasi-hydrostatic cooling). In their semi-analytical models, \citet{Merson2013} recalculated their predicted number counts in two different ways: one representing model $BzK$ galaxies with a simple K-mag limit brighter than K$_{s}$ $\leq$ 24 and a second one which also takes into account
a B-band detection limit of B $\leq$ 28.

The oldest model among those we examine \citep{Kitzbichler2007} fails to reproduce the number counts at all magnitudes, while the newer models \cite[]{Somerville2012, Merson2013} do better, although the \citet{Merson2013} model underpredicts counts at the bright end.  None of the models extend to the brightest passive galaxies that we observe (\Ks<19.5) and examining such an extension would provide an interesting test of the models, and in particular whether their AGN feedback prescription can account for the observed number counts or needs to be modified or amended by additional physics. 


\begin{table}
\label{tab:numberCounts}.
\caption[]{Number counts of \PEgzK\ and \SFgzK\ galaxies, corrected for detection incompleteness. For star-forming galaxies, numbers outside of the parenthesis represent number counts of only detected in g galaxies, whereas those inside the parenthesis represent an upper limit, including galaxies that were detected and not-detected in \g. } 
\begin{centering}
\begin{tabular}{|c|c|c|}
\hline 
$K_{s}$ & PE-gzK$_{s}$  & SF-gzK$_{s}$ \\
(AB) & {[}N/$deg^{2}/0.5mag${]} & {[}N/$deg^{2}/0.5mag${]} \\
\hline
15.25 & 0 & 0.04$\pm$0.04 (0.04$\pm$0.04) \\
15.75 & 0 & 0.2$\pm$0.1 ($0.2\pm$0.1) \\
16.25 & 0 & 0 (0) \\
16.75 & 0 & 0.2$\pm$0.1 (0.2$\pm$0.1) \\
17.25 & 0 & 0.58$\pm$0.14 (0.58$\pm$0.14) \\
17.75 & 0 & 0.87$\pm$0.18 (0.87$\pm$0.18) \\
18.25 & 0 & 2.4$\pm$0.3 (2.4$\pm$0.3) \\
18.75 & 0.2$\pm$0.1 & 5.07$\pm$0.43 (5.11$\pm$0.43) \\
19.25 & 2.1$\pm$0.3 & 13.5$\pm$0.7 (13.5$\pm$0.7) \\
19.75 & 19.5$\pm$0.8 & 43.1$\pm$1.2 (43.1$\pm$1.2) \\
20.25 & 87.3$\pm$1.8 & 176$\pm$3 (177$\pm$3) \\
20.75 & 238$\pm$10 & 477$\pm$14 (488$\pm$14) \\
21.25 & 399$\pm$13 & 1140$\pm$21 (1217$\pm$22) \\
21.75 & 509$\pm$14 & 2288$\pm$30 (2596$\pm$32) \\
22.25 & 462$\pm$14 & 4034$\pm$40 (4765$\pm$44) \\
22.75 & 345$\pm$12 & 5934$\pm$49 (7211$\pm$54) \\
23.25 & 312$\pm$14 & 7688$\pm$55 (9998$\pm$63) \\
\hline 
\end{tabular}
\par\end{centering}
\end{table}

 \begin{figure}
	\includegraphics[width=\columnwidth]{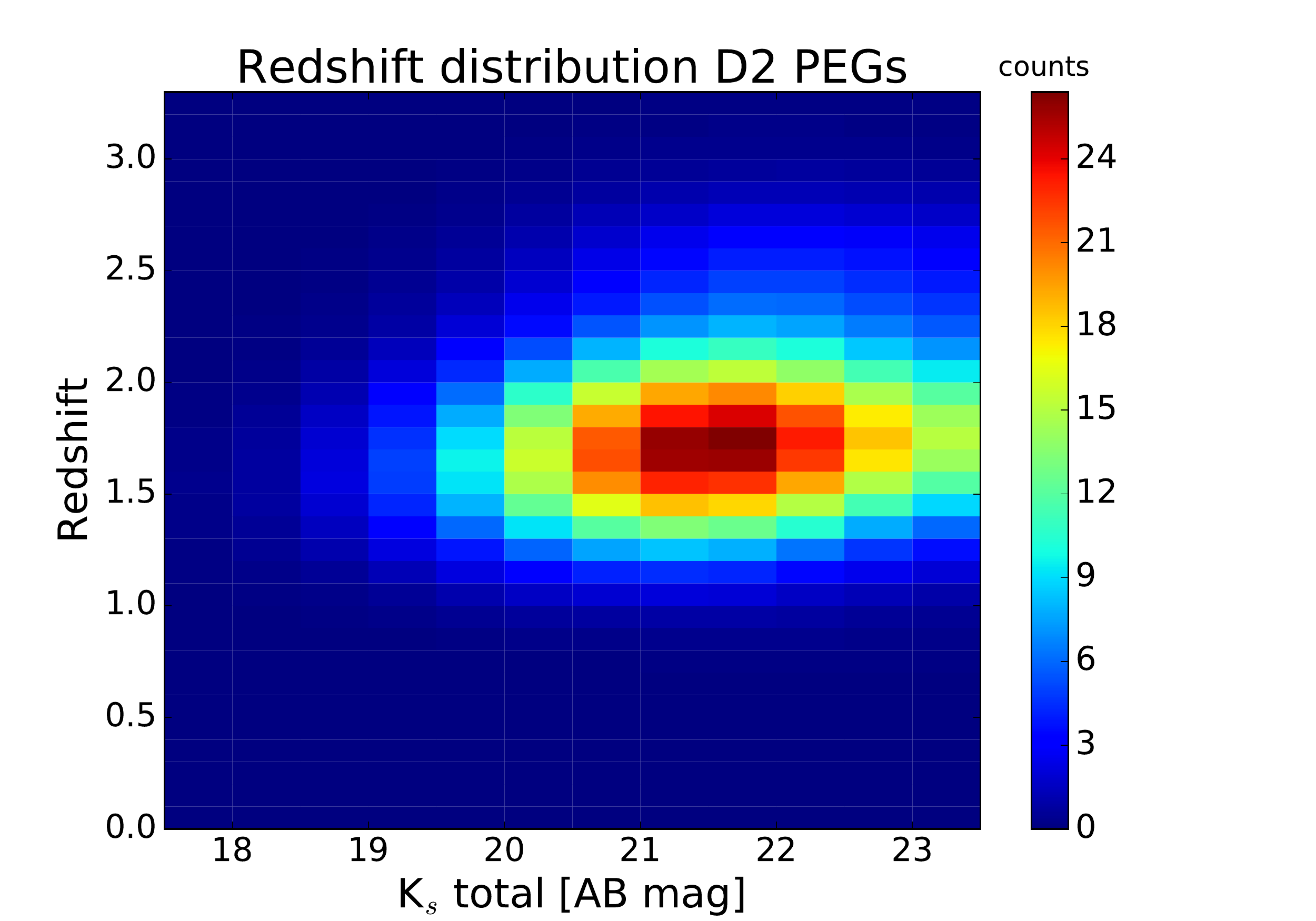}
	\includegraphics[width=\columnwidth]{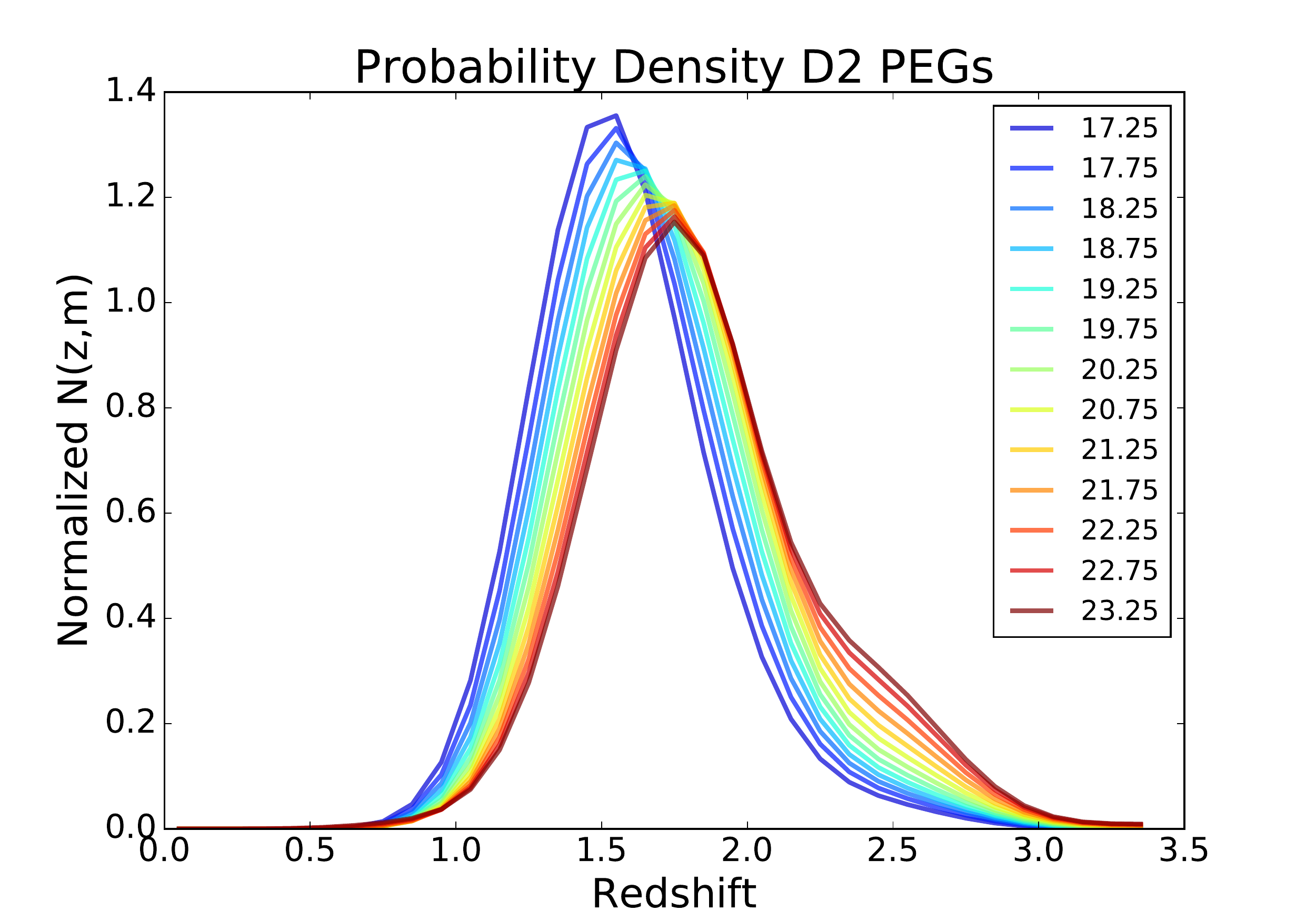}
    	\includegraphics[width=\columnwidth]{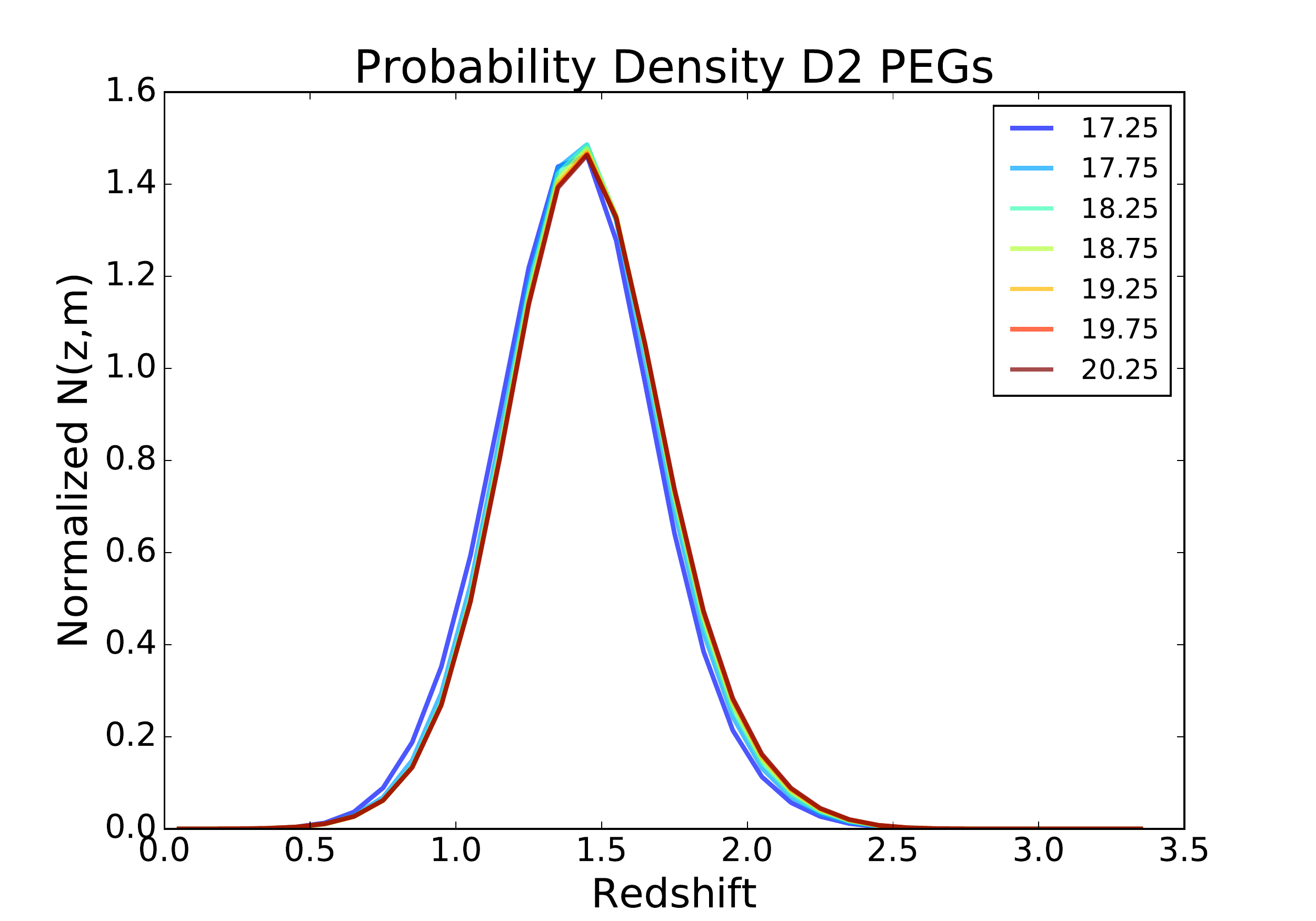}
    \caption{Top panel: number of \PEgzK\ galaxies as a function of magnitude and redshift, with redshifts taken from Muzzin et al.\ (2013). Middle panel: redshift distributions as a function of magnitude, constructed after mild smoothing of the data in the top panel. Bottom panel: similar to the middle panel, but using photometric redshifts from Moutard et al.\ (2016b). 
    }
    \label{fig:redshiftDistributions}
\end{figure}

\section{Stellar Mass Function}
\label{sec:SMF}

\subsection{Procedure}

We take a simple approach to estimating the stellar mass function of our \PEgzK\ galaxies:  we start with the number counts already detailed in Sec.~\ref{sec:numberCounts}; we then estimate the (magnitude-dependent) redshift distribution of the population using 30-band photometric redshifts for our \PEgzK\ galaxies in the D2 (COSMOS) field, which allows us to estimate effective volumes of our entire survey; and we convert \Ks-band magnitudes to stellar masses using an empirical relation derived, again, in our D2 (COSMOS) subsample.   These ingredients are described in detail below. 

\subsubsection{Redshift distribution and effective volumes}\label{sec:Veff}

Although our Wide fields contain $ugriz+K_s$ data from which \citet{Moutard2016b}  derived photometric redshifts, the lack of $J$ and $H$ bands in this dataset means that redshift estimates at redshifts of interest to us suffer from large degeneracies due to the inability to constrain the location of the Balmer/4000\AA\ breaks.  Therefore, instead of relying on photometric redshifts of individual objects, we take a statistical approach and use the photometric redshift distribution for a subsample of our objects where the photometric redshifts are likely to be very good, namely in the D2 (COSMOS) field.  We cross-matched our D2 \PEgzK\ galaxies with the \citet{Muzzin2013cat} 30-band COSMOS photo-$z$ catalogue, resulting in 1384 \PEgzK\ objects with excellent photo-z's.  The top panel of Fig~\ref{fig:redshiftDistributions} shows the result and it is clear that the redshift distribution is magnitude-dependent.  

We smoothed our 2D histogram at each magnitude bin using a Gaussian kernel, which extrapolates at both edges of the distribution, i.e, the brightest and faintest bins, and we normalized the result by the area under the curve to obtain a redshift probability density distribution, shown in the middle panel of Fig.~\ref{fig:redshiftDistributions}. The redshift distribution peaks at $z\sim1.6$ for bright \PEgzK\ galaxies and at somewhat higher redshifts for fainter ones; in all cases there is a non-negligible non-Gaussian tail to higher redshifts.  For comparison, in the bottom panel of Fig.~\ref{fig:redshiftDistributions} we also show the redshift PDFs derived in a similar way, but using the photometric redshifts of \cite{Moutard2016b}. We adopt the PDFs based on the \cite{Muzzin2013cat} photometric redshifts for the rest of this paper. 

We note that the initial assessment of the $BzK$ galaxy population redshift distribution by \citet{Daddi2004b} was based on a much smaller sample with a limited number of bright galaxies. Since most of their sample lies at fainter magnitudes, they reported a redshift distribution of $1.4 < z < 2.5$. With the advent of larger area surveys with good photometric redshifts we can see that the redshift distribution peaks at the lower end of this redshift range, particularly for the brighter objects. 

We assume that the normalised redshift density distributions shown in the middle panel of Fig.~\ref{fig:redshiftDistributions} are a good proxy for our redshift probability density function, or $p(m,z)$, which we will use later in estimating the effective volume of the survey. Starting from the $p(m,z)$ we can then estimate the magnitude-dependent effective volumes $V_{eff}(m)$ for our sample. Following \citet{Steidel1999} and \citet{Sawicki2006}, this is done using
\begin{equation}\label{eq:veff}
V_{eff}(m)=A\int\frac{dV}{dz} p(m,z) dz, 
\end{equation}
where A is the effective area of our survey (27.60 \sqdeg\ for galaxies with \Ks < 20.5 and 2.51 \sqdeg\ for fainter galaxies), $dV/dz$ is the comoving volume per square degree and $p(m,z)$ is the redshift- and magnitude-dependent PDF that we estimated from the COSMOS photometric redshift analysis.  Note that this approach is somewhat different from that of \citeauthor{Steidel1998} and \citeauthor{Sawicki2006}, since those authors assumed that not all the LBGs in their surveys were detected, whereas we assume that all of our \PEgzK\ galaxies are.

\subsubsection{Estimating masses from \Ks\ magnitudes}

 \begin{figure}
	\includegraphics[width=\columnwidth]{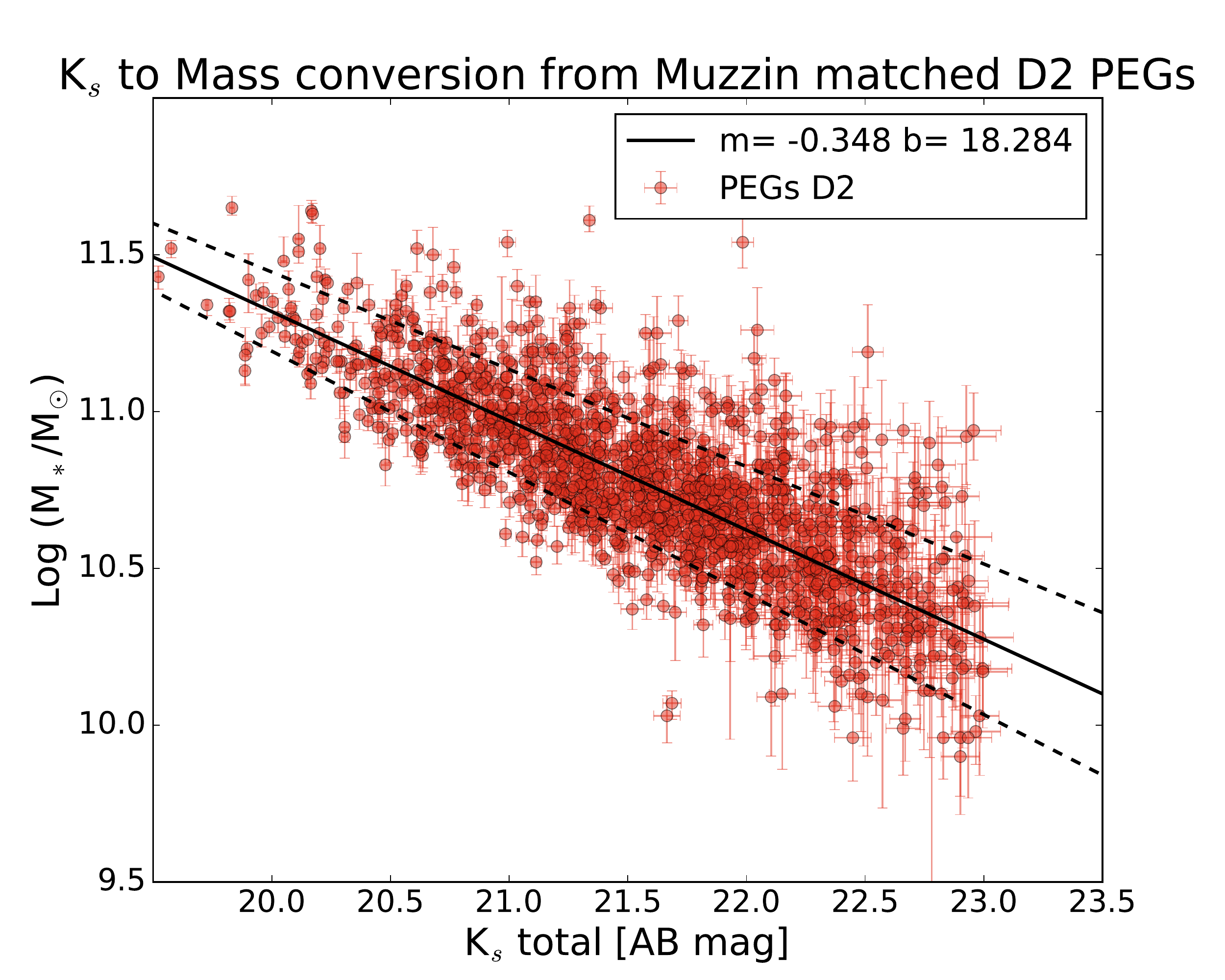}
    \caption{Correlation between PE-\gzK\ galaxy \Ks\ magnitudes and stellar mass. Stellar masses were obtained from the catalogue of \citet{Muzzin2013cat} after cross-matching $\sim$1400 objects in D2 (COSMOS). The best fit correlation is shown as a solid line and its 1$\sigma$ scatter is represented with dashed lines. At $K_s \sim 20.5$ this $1\sigma$ scatter is approximately 0.12 in $\log(M_{stars}/M_\odot)$. As can be seen in this figure, there seems to be a good correlation between $K_s$ band magnitudes and the stellar mass of a PE galaxy.}
    \label{fig:KtoMass}
\end{figure}

We estimated galaxy masses for our passive galaxies from their \K\ total magnitudes. At redshifts of z $\sim$ 1.6, the peak of the redshift distribution of luminous \PEgzK\ galaxies, the observed \K\ band corresponds to rest-frame 9200\AA\ and at these wavelengths the light from these galaxies is generated by the long-lived low-mass stars that are expected to contain most of the stellar mass of the galaxy.

To find a suitable conversion between magnitude and mass we used the same $\sim$1400 cross-matched \PEgzK\ galaxies in D2 (COSMOS) that we used to work out the redshift distribution.  To estimate galaxy stellar masses  \citet{Muzzin2013cat} followed the now standard SED-fitting approach first introduced by \cite{Sawicki1998} in  which multi-wavelength broadband photometry is compared with a grid of predictions derived from redshifted and dust-attenuated stellar population synthesis models.  Here, \citet{Muzzin2013cat} used a set of models with exponentially declining star-formation histories (SFHs), assumed solar metallicity, the \citet{Chabrier2003a} IMF, and allowed visual attenuation (A$_{V}$) to vary between 0 and 4.  As expected for passive galaxies, the masses scale well with observed \Ks-band total magnitudes as shown in Fig.~\ref{fig:KtoMass}. The best-fit orthogonal distance correlation is shown as a solid line and is described by 
\begin{equation}
log[M_\star/M_{\odot}]=-0.348K_{s}+18.284
\label{eq:ks_mass}.
\end{equation}
We use this relation to convert the observed magnitudes of our galaxies to stellar masses.

\subsection{Results}

Using our empirical magnitude-to-mass relation (Eq.~\ref{eq:ks_mass}), we estimated stellar masses for all our \PEgzK\ galaxies from their \Ks\ magnitudes. We then counted  \PEgzK\ galaxy numbers as a function of stellar mass in logarithmically-spaced mass bins.  Next, we corrected these numbers for incompleteness using the same approach as we used for the number counts in \S~\ref{sec:numberCounts} but after first using Eq.\ ~\ref{eq:ks_mass} to convert our magnitude-incompleteness estimates of \S~\ref{sec:numberCounts} into incompleteness as a function of mass.  Finally, we divided the resulting numbers by the effective volumes from \S~\ref{sec:Veff} after using Eq.~\ref{eq:ks_mass} once again to convert from magnitude-dependent effective volumes into mass-dependent effective volumes.  The result is the stellar mass function shown with red points and connecting dashed red line in Fig.~\ref{fig:SMFcompilation}.

\subsubsection{Direct comparison with previous work}

 \begin{figure}
	\includegraphics[width=\columnwidth]{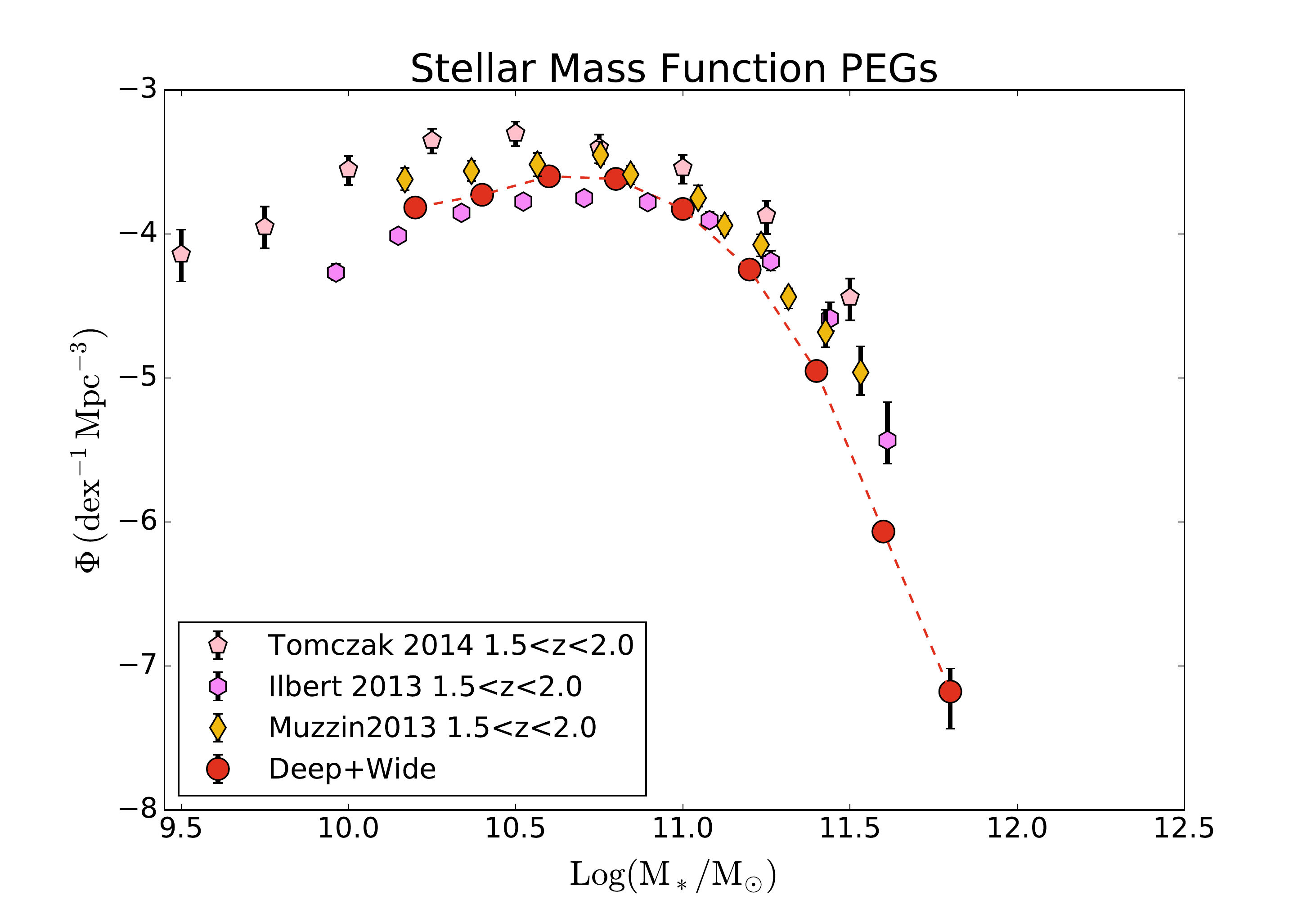}
    \caption{Stellar Mass Functions of passive galaxies at $z\sim 1.6$.  Our results (red points and connecting line) are consistent with previous studies at intermediate masses but are lower at high masses. Note that the much larger area of our survey allows us to constrain the massive end of the SMF with higher significance than was done previously. }
    \label{fig:SMFcompilation}
\end{figure}

Using the photometric redshift distribution along with the \Ks-\Mstars\ relation (i.e., Eq.~\ref{eq:ks_mass}) we built a \zs1.6 passive galaxy stellar mass function with our combined Deep and Wide catalogues. Our direct SMF measurement is shown with red circles in Fig.~\ref{fig:SMFcompilation}, along with the PE galaxy SMFs of \citet{Muzzin2013smf}, \citet{Ilbert2013b}, and \citet{Tomczak2014} at similar redshifts. 

As can be seen in Fig.~\ref{fig:SMFcompilation}, there are significant differences among these previous studies, all of which are based on relatively small areas:  0.5~deg$^{2}$ for \citet{Tomczak2014} and 1.6~deg$^{2}$ for \cite{Muzzin2013smf} and \cite{Ilbert2013b} (both covering the same area in the COSMOS field). While cosmic variance can be expected to play a role in such relatively small fields, the observed scatter between surveys is not likely to be due to cosmic variance alone since the \citet{Muzzin2013smf} and \citet{Ilbert2013b} results, which differ significantly, are based on independent analyses of the same dataset in the COSMOS field. Instead, sample selection or other systematics may play a significant role in producing the discrepancies.  

As Fig.~\ref{fig:SMFcompilation} shows, our results are consistent with previous studies at intermediate and low masses (\Mstars$\la$10$^{11}$\Msun), lying within the number density range spanned by these previous works. However, at larger masses, \Mstars $\ga$10$^{11}$\Msun, we measure a lower number density than the previous studies. We note that our statistics at the massive end, based on a 27.6~\sqdeg\ dataset, are significantly better than any previous work. Our points shown in Fig.~\ref{fig:SMFcompilation} have not been corrected for Eddington bias \citep{Eddington1913}; we will correct for this in \S~\ref{sec:schechterFit}\footnote{Note that only some authors correct for Eddington bias: while \citep{Ilbert2013b} do, it appears that \citep{Muzzin2013smf} and \citep{Tomczak2014} do not.} .  As noted in \S~\ref{sec:schechterFit}, correcting for Eddington bias brings our number density down somewhat at the massive end, exacerbating the discrepancy with the previous studies. 

The discrepancy between our measurement and previous work may be due to sample variance or cosmic variance dominating the smaller fields (we have more than an order of magnitude more area than the earlier studies and, as noted earlier, the results of \cite{Muzzin2013smf} and \cite{Ilbert2013b} are not independent as they are based on a single field). Alternatively, systematic differences in sample selection, assumptions about objects' redshift distributions, or mass measurement may also play a role. Full resolution of this issue will require spectroscopy of a significant number of the massive galaxies in our sample. In the meantime, given that at intermediate and low masses our measurements are consistent with previous work, we feel it plausible that at the massive end our measurement is likely more correct than the previous work based on two relatively small fields.

\subsubsection{Schechter function fitting and Eddington bias correction}\label{sec:schechterFit}

 \begin{figure}
	\includegraphics[width=\columnwidth]{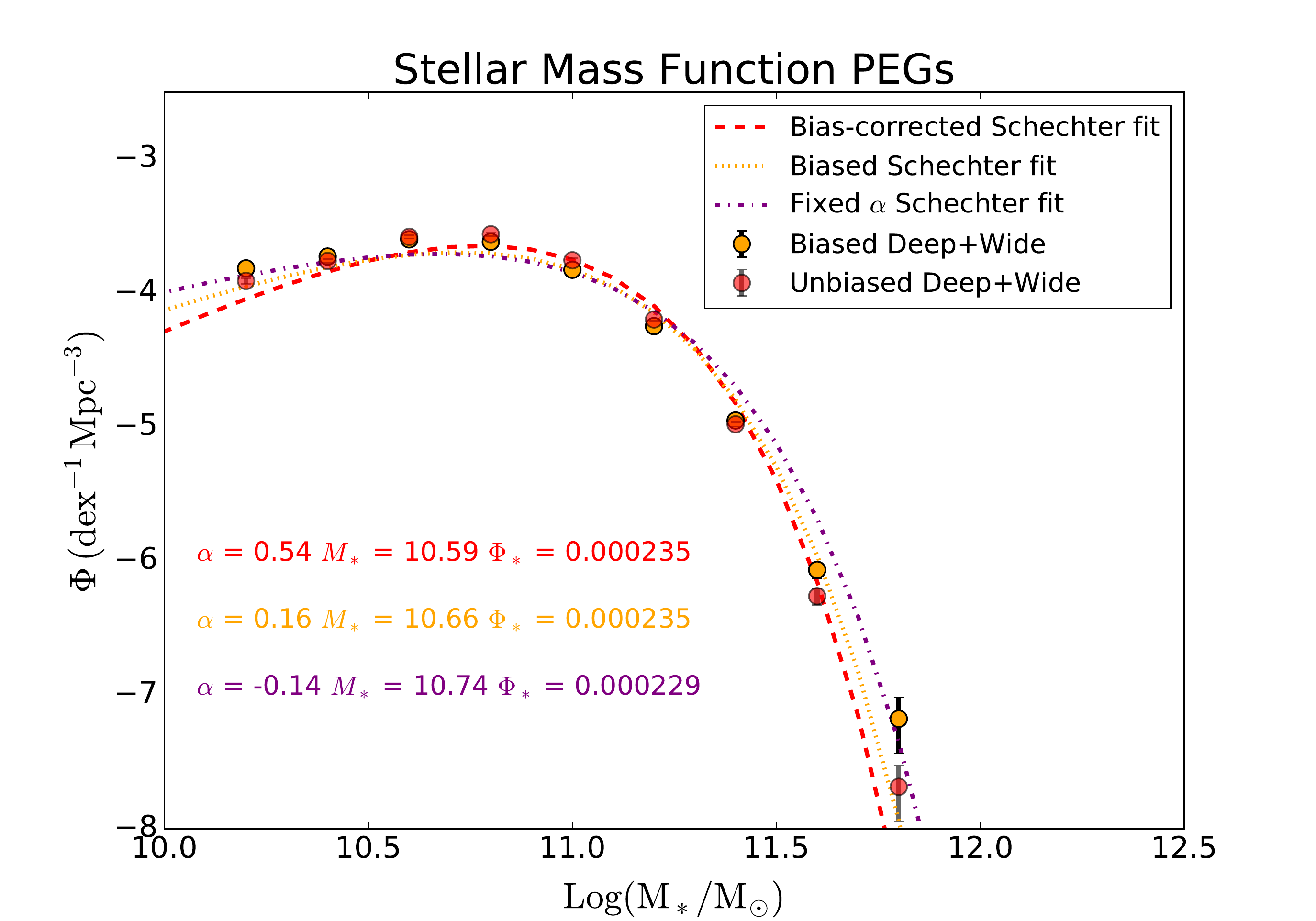}
    \caption{Best-fit Schechter function for our \PEgzK\ galaxies. The red dashed line represents the best-fit Schechter function once it has been corrected for Eddington Bias. For comparison, using a dotted orange line we show the best-fit Schechter function when this effect is not taken into account. Using the fractional change between the biased and bias-corrected Schechter fits, we correct the observed stellar mass function data: Orange represents the observed, biased SMF while red represent the Eddington-bias corrected SMF.}
    \label{fig:SMFfinal}
\end{figure}

 \begin{figure}
	\includegraphics[width=9cm]{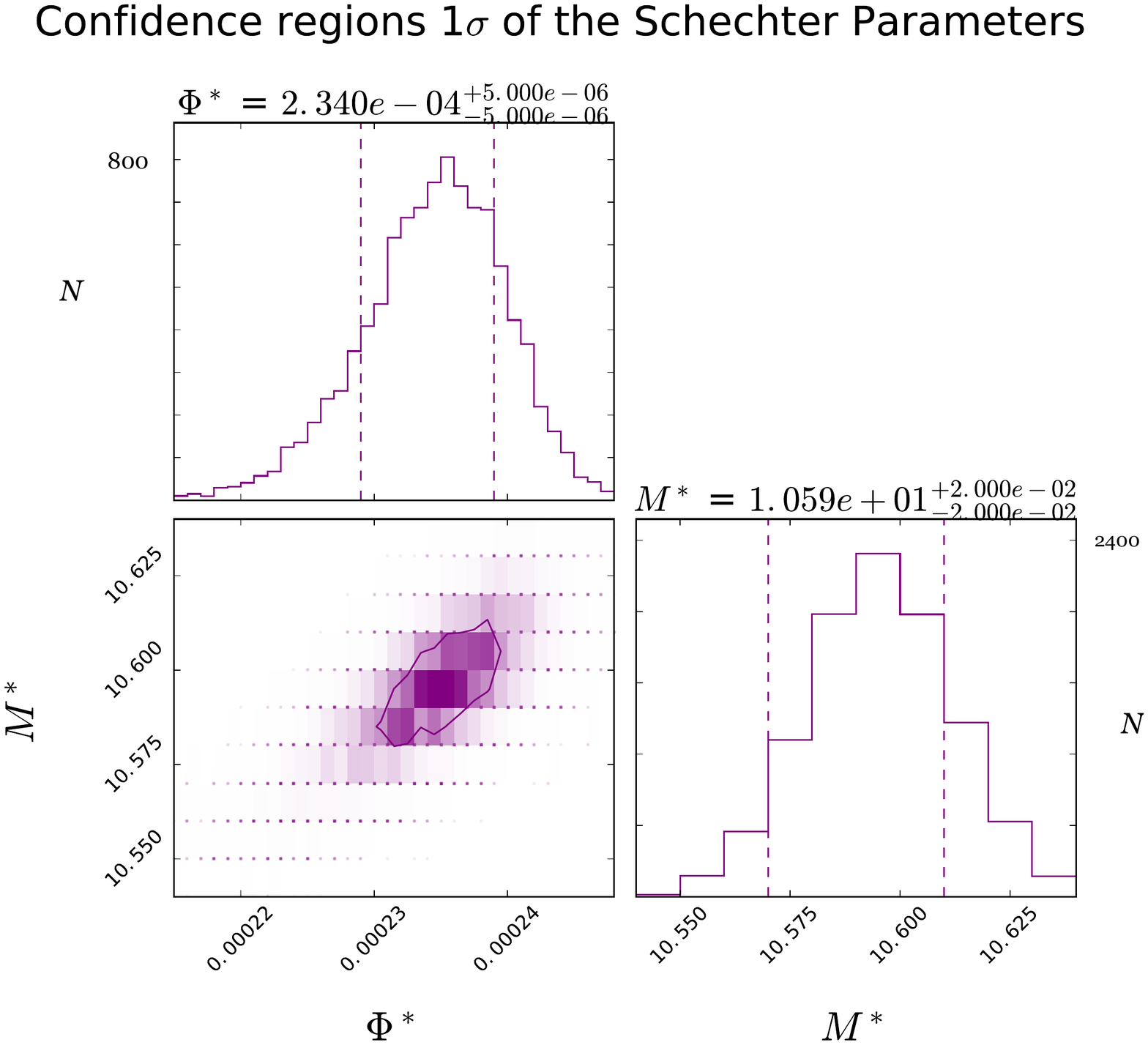}
    	\includegraphics[width=9cm]{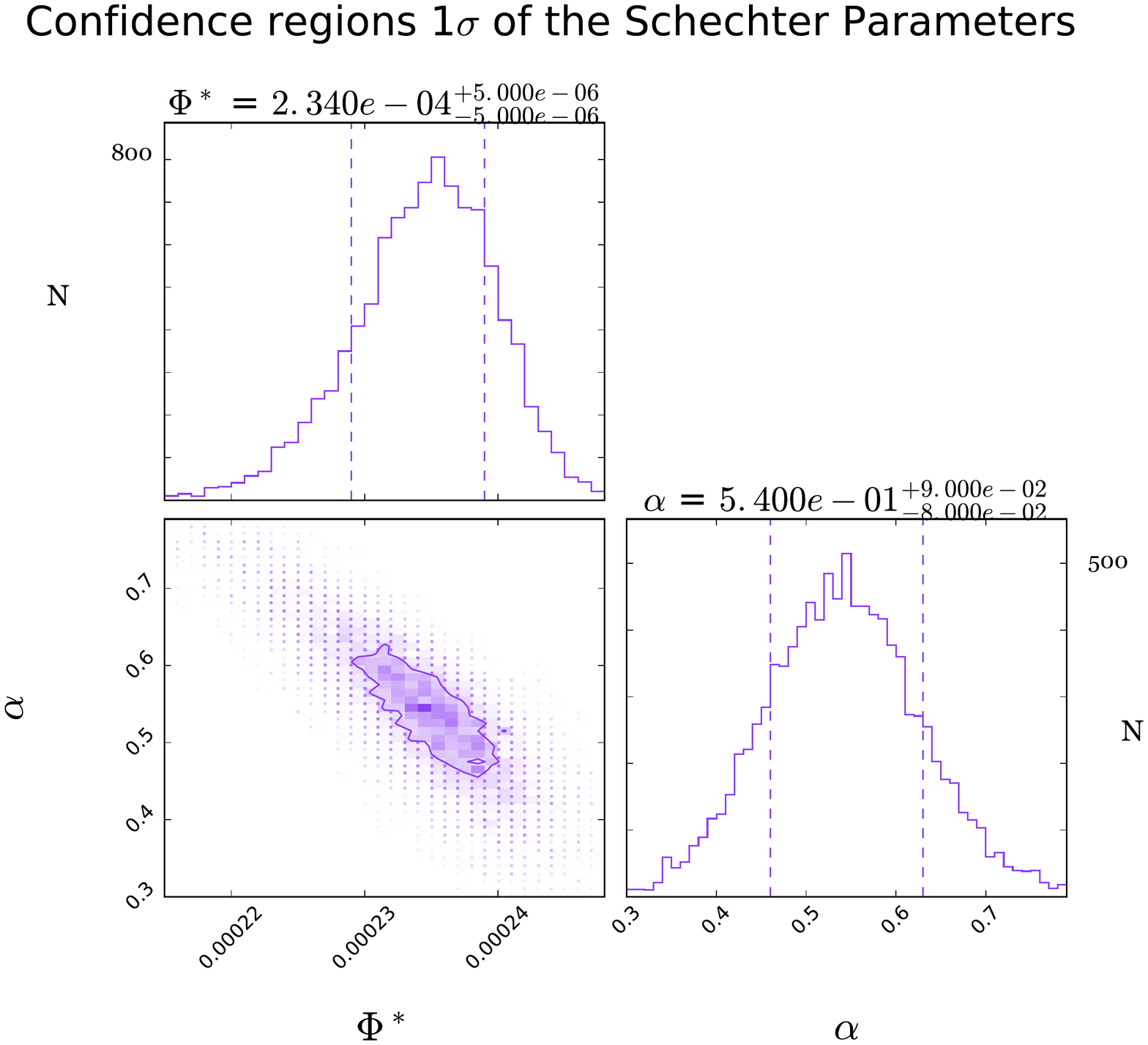}
    \caption{ One sigma confidence regions for the best-fit Schechter parameters after 10000 bootstrap resamplings of our passively evolving\gzK\ galaxies. Labels at the top of each histogram show the median for each parameter.}
    \label{fig:SchechterParameters}
\end{figure}

The Schechter function \citep{Schechter1976} is a fitting formula that, with just three adjustable parameters, is found to well describe the galaxy luminosity and stellar mass functions at many epochs \citep[e.g.,][]{Lilly1995, Sawicki1997, Sawicki2006, Peng2010, Ilbert2013b, Muzzin2013smf, Moutard2016b}. The functional Schechter form can be physically motivated, but we leave such aspects to \S~\ref{sec:discussion} and in the present section focus on its purely operational role as a fundamental description of a galaxy population.

To find the Schechter function that best describes our population of passively evolving \PEgzK\ galaxies, we performed a $\chi^{2}$ minimisation fit to our mass function data. However, since galaxy density decreases exponentially towards brighter magnitudes, in the presence of uncertainties in mass estimates, galaxies are more likely to scatter to higher than to lower masses. This effect, called the Eddington Bias \citep{Eddington1913} can cause an apparent steepening of the stellar mass function at the massive end. To assess and correct for the effect of this bias, we also performed a $\chi^{2}$ fit on a Schechter function convolved with typical stellar mass errors.

To begin, we bootstrap resample our original sample, then perturb the masses in the  bootstrapped catalogue using a normal distribution with mass-dependent width that characterises the scatter seen in Fig.~\ref{fig:KtoMass}. By performing the bootstrap and mass perturbation on our sample we account for sample variance and mass uncertainties, respectively, when calculating the best-fit model.  

\begin{table*}
\begin{centering}
\begin{tabular}{|l|c|c|c|c|c|c|}
\hline 
Survey &  Field(s) & area[\sqdeg] & selection & log($M^{*})[M_{\odot}]$ &$\Phi^{*}[10^{-4}Mpc^{-3}]$ & $\alpha$\\
\hline 
LARgE (this work), free $\alpha$ & CFHTLS W+D & 25.1+2.5 & PE-\gzK &  10.59 $\pm$ 0.02 & 2.35 $\pm$ 0.05 & 0.54 $\pm$ 0.08\\
LARgE (this work), fixed $\alpha=-0.14$ & CFHTLS W+D & 25.1+2.5 & PE-\gzK & 10.74 $\pm$ 0.01 & 2.29 $\pm$ 0.03 & $-$0.14\\
\cite{Muzzin2013smf}, free $\alpha$ & COSMOS & 1.6 & photo-z + SED & $10.67\pm0.03$ & $4.15 _{-0.08}^{+0.06}$ & 0.03 $\pm$ 0.11\\
\cite{Muzzin2013smf}, fixed $\alpha=-0.4$ & COSMOS & 1.6 & photo-z + SED & $10.80\pm0.01$ & $3.61 _{-0.04}^{+0.02}$ & -0.4\\
\cite{Ilbert2013b} & COSMOS & 1.6 & photo-z + SED &  10.73 $_{-0.04}^{+0.03}$  & 2.20 $_{-0.01}^{+0.01}$  & 0.10$ _{-0.09}^{+0.09}$ \\
\cite{Tomczak2014} & NEWFIRM+ZFOURGE & 0.4+0.1 & photo-z + SED & 10.76 $\pm$ 0.05 & 3.29 $\pm$ 0.05 & -0.14 $\pm$ 0.12\\
\hline 
\end{tabular}
\par\end{centering}
\protect\caption[Compilation of best-fit Schechter function parameters.]{Best fit Schechter function parameters for quiescent galaxies at $1.5\leq$z$\leq2.0$.  When two values are listed under area then one refers to the wide and the other to the deep component of a "wedding cake" analysis.  We present two results for our sample: one with a free $\alpha$ and a fixed $\alpha = -0.14$ (the value given in \cite{Tomczak2014}). Similarly, \cite{Muzzin2013smf} have a free-$\alpha$ and a fixed $\alpha = -0.4$ result. 
\label{tab:best-fit-params} }
\end{table*}

Next, for each bootstrap, we bin our data and perform a $\chi^{2}$
minimisation using the Schechter function 
\begin{equation}
\Phi(M)=ln\left(10\right)\Phi^{*}\left[10^{\left(M-M^{*}\right)\left(1+\alpha\right)}\right]exp\left[-10^{(M-M^{*})}\right],
\label{eq:sch}
\end{equation}
\cite[]{Schechter1976, Marchesini2009}
that has been convolved with a Gaussian using 
\begin{equation}
\Phi_{c\!o\!n\!v\!o\!l\!v\!e\!d}(M)=\intop_{-\infty}^{\infty}\Phi(M')G(M-M',\sigma)dM',
\label{eq:sch_conv}
\end{equation}
where 
\begin{equation}\label{eq:SchConvKernel}
G(M-M', \sigma)=\frac{1}{\sigma\sqrt{2\pi}}\:\exp\left(\frac{M-M'}{2\sigma^{2}}\right)
\end{equation}
represents the scatter in the \Ks-\Mstars\ relation.  Because photometric uncertainties in the \Ks-band are very small for the bright objects that populate the steep part of the SMF, the dominant source of Eddington Bias is the scatter in the \Mstars-\Ks\ relation and this is taken into account by the convolution described by Eqs.\ ~\ref{eq:sch_conv} and \ref{eq:SchConvKernel}.

The free parameters in the fit are  $\alpha$,
$M^{*}$ and $\Phi^{*}$, as defined by the Schechter function (Eq.~\ref{eq:sch}). We performed 10,000 bootstrap resamples and carried out parameter minimisation on each realisation of the dataset. We define our best-fit model to be the peak of our output best-fit parameter space.  The best-fit Schechter function, after correcting for Eddington bias, is shown as the red line in Fig.~\ref{fig:SMFfinal}; the orange line shows the Schechter fit without the Eddington bias correction.   We also show the original, uncorrected data (orange points) and, in red, data points corrected for Eddington bias by applying the ratio of the two (corrected and uncorrected) Schechter functions. The probability distributions of the Schecher parameters, based on the 10,000 resampled fits, are shown in Fig.~\ref{fig:SchechterParameters}.

In Table~\ref{tab:best-fit-params} we summarise our best-fit Schechter parameters, along with the results of three other recent studies \citep{Ilbert2013b, Muzzin2013smf, Tomczak2014}.  Comparing the different results, we see that the values of $M^*$ cluster closely together, although are often formally inconsistent with one another given the small uncertainties. The values of characteristic density, $\Phi^*$ vary significantly between the surveys, in part, though probably not wholly, reflecting differences between PE galaxy selection criteria and --- for the smaller-area surveys --- cosmic variance. Finally, there is a large scatter in the values of the slope at the low-mass end, $\alpha$, along with large associated uncertainties. This is not unexpected, given that in order to constrain this low-mass end slope it is necessary to go deep below $\log (M^* / M_\odot ) < 10.5$. 

Our own Schechter fit is strongly influenced by the high-S/N points around $M^{*}$ where galaxies are most numerous intrinsically and where we also have excellent statistics from the Wide fields. Figure~\ref{fig:SMFcompilation} shows that among the surveys considered only the sample of \cite{Tomczak2014} reaches deep into the low-mass end (albeit with poor statistics) and thus likely gives the best constraint on $\alpha$. For this reason we also re-fit our data fixing $\alpha$ to the $-0.14$ of \citeauthor{Tomczak2014} Fixing $\alpha$ in this way changes our $M^*$ and $\Phi^*$ somewhat and the resulting Schechter parameters are also shown in Table~\ref{tab:best-fit-params} and illustrated with the dotted black line in Fig.~\ref{fig:SMFfinal}.

\section{Discussion}\label{sec:discussion}

\subsection{The growth of the quenched galaxy population}

 \begin{figure}
	\includegraphics[width=\columnwidth]{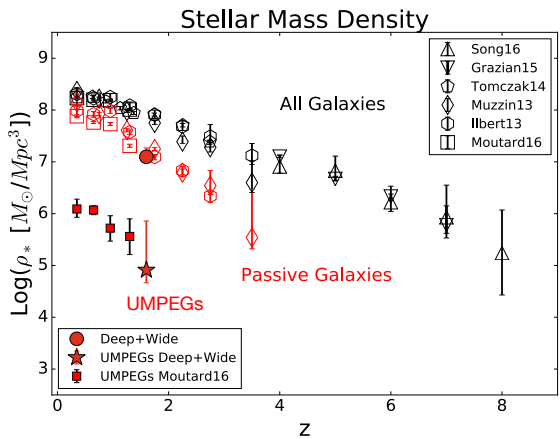}
    \caption{Cosmic stellar mass density as a function of redshift. Results from previous studies are shown with open symbols, while our result for passively evolving $gzK_s$ galaxies in the range $8 < \log$(\Mstars/\Msun)$< 13$ is shown as a red filled diamond, while ultra massive passive galaxies ($11.5 < \log$(\Mstars/\Msun) $< 13$) are represented with the filled red star. 
    }
    \label{fig:SMD}
\end{figure}

It is interesting to investigate how the quiescent galaxy population grows in time.  We first examine the growth of the comoving stellar mass density, shown in Fig.~\ref{fig:SMD}.  Given that we probe the peak of the PE galaxy SMF with excellent statistics derived from our large area, we are in an excellent position to constrain the global stellar mass density contained in passive galaxies at \zs1.6. 

In Fig.~\ref{fig:SMD} red points show the growth measured by several studies of the stellar mass density due to passive galaxies with $8<\log$(\Mstars/\Msun) $<13$ (except for \citet{Tomczak2014} who integrated down to $\log$(\Mstars/\Msun)$=$8). All points have been changed, when necessary, to the \cite{Chabrier2003a} IMF.  Black points show the growth of the SMD due to galaxies of all types while the red points show the contribution of the quiescent galaxy population. Our \PEgzK\ galaxies' contribution at \zs1.6, shown with the filled red circle is calculated by integrating our PE galaxy Schechter function over $8<\log$(\Mstars/\Msun) $< 13$ and is consistent with previous studies.  Overall, as is well known, the stellar mass density in passive galaxies grows more rapidly than the total stellar mass density, reflecting the growing importance of the quenched population with cosmic time. 

In this project we are particularly interested in the extremely rare, ultra-massive passive galaxies and so we also calculate the stellar mass density of \PEgzK\ galaxies with $11.5<\log$(\Mstars/\Msun)~$<13$ and show it with the red star in Fig.~\ref{fig:SMD}.  We also show (filled red squares) the UMPEG contribution to the SMD at lower redshifts obtained by integrating the quiescent galaxy SMFs of \cite{Moutard2016b}. While individually very massive, UMPEGs are also very rare, and so, as a population, do not contribute significantly to the total stellar mass density:  they account for only $\sim0.66\%$ of the mass contained in quiescent galaxies at \zs1.6 and a similar fraction at lower redshifts. These are rare monsters indeed. 

 \begin{figure}
	\includegraphics[width=\columnwidth]{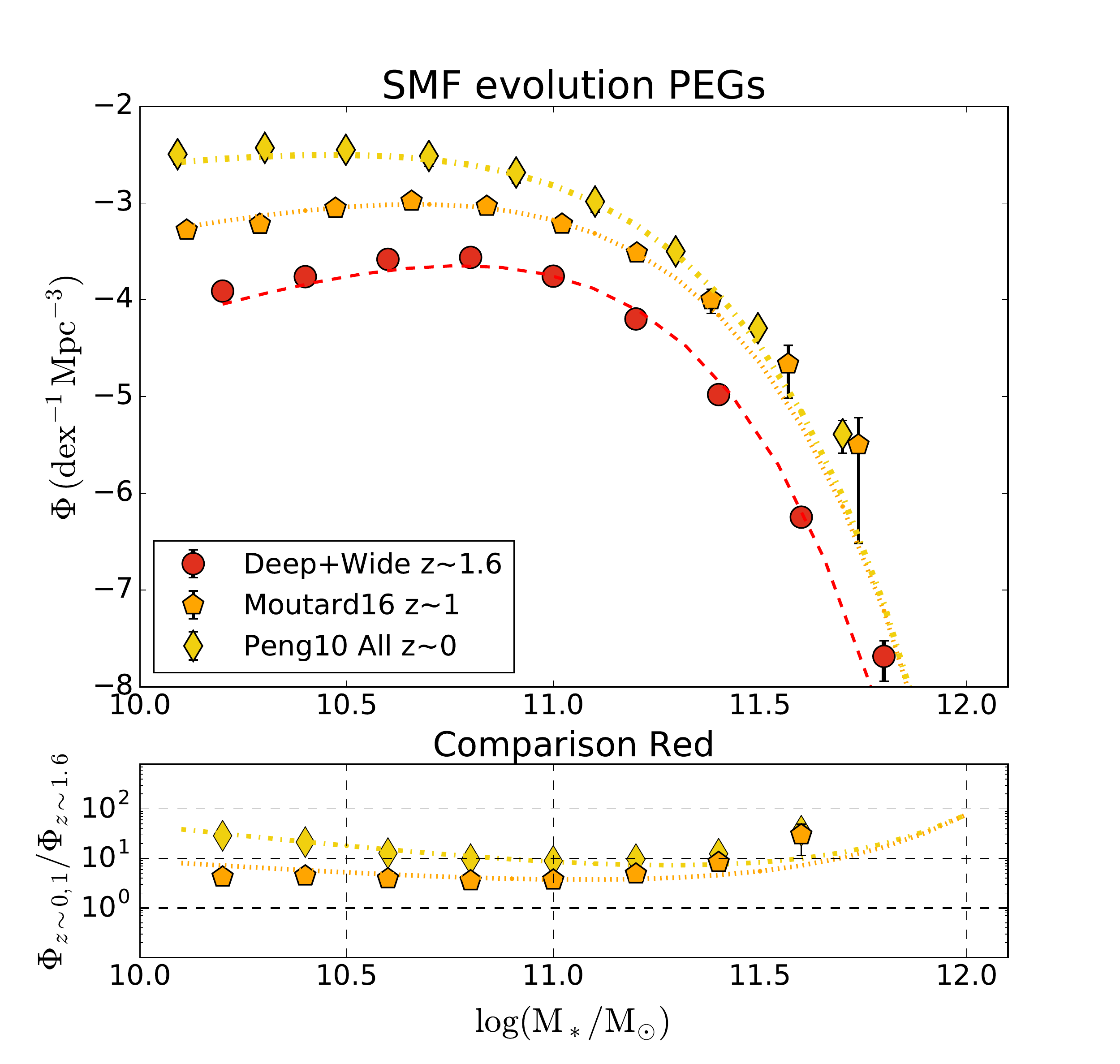}
    \caption{ Top panel: Stellar Mass Function of passive galaxies and their best-fit Schechter models.  Bottom panel: ratios of the SMFs.   As can be seen from this panel, the evolution of these $z\sim1.6$ galaxies compared to $z\sim1$ and $z\sim0$ is relatively smooth for intermediate masses. However, at the faint end, where environmental quenching is expected to become more important at lower redshifts, we see strong growth in the low-mass population.}
    \label{fig:SMFevolution}
\end{figure}

We next turn to the evolution of the passive galaxy stellar mass function, which is shown in Fig.~\ref{fig:SMFevolution}. In the top panel of this Figure we show our \zs1.6 quiescent galaxy stellar mass function (corrected for Eddington bias, \S~\ref{sec:schechterFit}) along with PE SMFs at \zs1 \citep{Moutard2016b} and \zs0 \citep{Peng2010}. The bottom panel shows the fractional change of the number density from \zs1.6 to \zs1 (orange pentagons) and from \zs1.6 to \zs0 (yellow diamonds). These ratios were calculated by interpolating between the points (and uncertainties) from each survey to the same log masses. Also shown, as lines, are the the ratios between the best-fit Schechter functions over the same redshift intervals. 

Overall, the quiescent galaxy population can be expected to grow in numbers as galaxies quench their star formation and migrate from the star-forming population to the quiescent one \citep[see, e.g., ][]{Peng2010}. The rates at which different parts of the quiescent galaxy SMF grow will then depend on the quenching mechanism(s) and other processes that post-process quenched galaxies, such as merging. With this in mind, it is interesting to examine how the SMF evolves over time. As can be seen in the bottom panel of Fig.~\ref{fig:SMFevolution}, the evolution of the PE galaxy SMF cannot be described as a uniform number density increase with time. Compared to the growth at intermediate masses, there is excess growth at the low-mass end, particularly from \zs1 to \zs0; there may also be excess growth at the high-mass end, albeit within large uncertainties.  The excess growth at low masses can be associated with the increasing importance of the environmental quenching mechanism  (a mass-independent quenching process, presumably associated with satellite quenching; \citealt{Peng2010}) which is expected to make an increasingly important contribution at later cosmic epochs as large scale structure develops in the Universe. In this context, it is not unexpected to see little excess growth in the low-mass end from \zs1.6 to \zs1, but then stronger growth down to \zs0. 

At the massive end it is hard to be sure of the differential growth between the high-mass and intermediate-mass parts of the SMF given the large uncertainties in the SMFs at lower redshifts. However, if the growth at the massive end is in excess of that at intermediate masses, it seems to be already happening between  $z\sim1.6$ and $z\sim1$. Such excess growth of the SMF could mean that during these later epochs quenching galaxies are significantly more massive than those quenched at higher redshifts, or that there is significant post-quenching mass growth through, for example, ("dry") mergers.

\subsection{The quenching of star formation in massive galaxies}

 \begin{figure}
	\includegraphics[width=8cm,height=7cm]{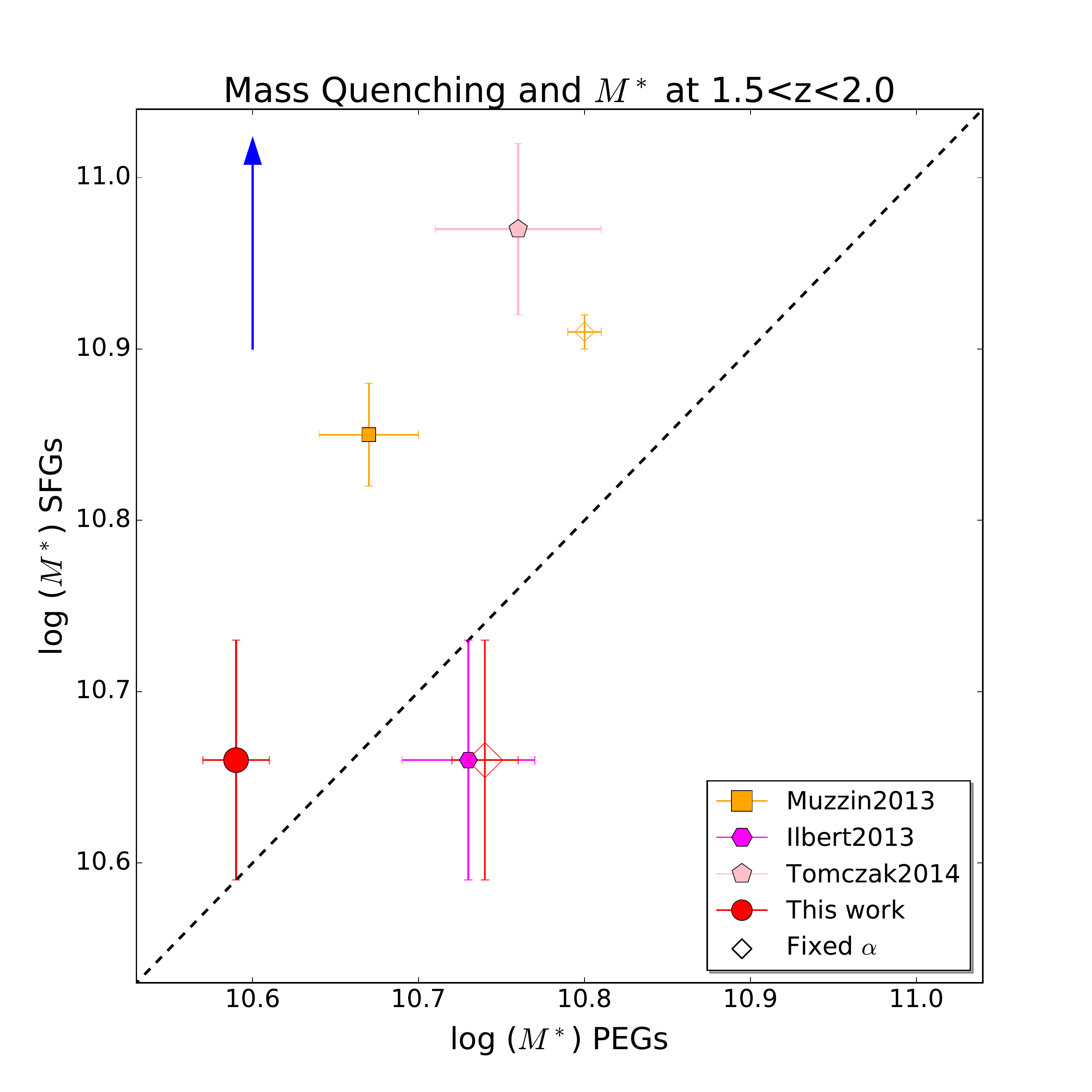}
    \caption{Comparison of best-fit characteristic mass $M^*$ of star-forming (SFG) vs.\ passive galaxies (PEG) at $1.5 < z < 2.0$. As before, red symbols represent this work (our best-fit $M^*$ for PEGs combined with Ilbert et al. best-fit $M^*$ for SFGs) with two best fits: one with alpha as a free parameter and a second fixing alpha to Tomczak's et al. best fit slope (i.e., $\alpha =  0.14$). Similarly, in orange, we show Muzzin et al. best-fit $M^*$ with a free and fixed slope (refer to Table~\ref{tab:best-fit-params}). Finally, Ilbert et al. and Tomczak et al. are represented in magenta and pink respectively. For most of these results, we observe that the best fit $M^*$ of SFGs is systematically higher ($\sim$ 0.2 dex mass) than the expected 1:1 correlation (shown with dashed black line) from Peng et al. Combining our results with Ilbert et al. is the only way our results would be in agreement (within uncertainties) with this 1:1 correlation. 
}
    \label{fig:MstarComparison}
\end{figure}

The Schechter function is not just an empirical fitting formula that has been found to reasonably describe the galaxy SMF: its functional shape can be physically motivated.  For example, \cite{Peng2010} postulated that the Shechter shape, both for star-forming and passive galaxy SMFs, is exactly produced by a ``mass quenching mechanism" that is of unknown physical origin but in which probability that a star-forming galaxy will quench and join the quiescent population increases exponentially with that galaxy's stellar mass. A separate ``environmental quenching mechanism'', independent of a galaxy's mass, is responsible for quenching star formation in lower-mass galaxies. The combination of these two mechanisms is found to be remarkably good at reproducing the observed shapes of the SMFs of both star-forming and quiescent galaxies at \zs0 \citep{Peng2010}. 

The effect of the environmental quenching mechanism is expected to grow with time as large scale structure develops in the Universe, and consequently it is minimised at high redshift. This means that at high redshift we can observe the mass quenching mechanism in isolation, uncontaminated by the admixture of environmentally-quenched galaxies.  In this context, the mass quenching mechanism predicts that the SMF of PE galaxies should be described by a single Schechter function of the form of Eq.~\ref{eq:sch}; it should be related to the SMF of SF galaxies, itself also a single Schechter function, through identical characteristic masses (i.e., $M^*_{PE} = M^*_{SF}$) and low-mass slopes that differ by $\alpha_{PE}-\alpha_{SF} \sim 1$ \citep{Peng2010}.

In \S~\ref{sec:SMF} (Fig.~\ref{fig:SMFfinal}) we presented the SMF of \zs1.6 PE galaxies. Using the statistics made possible by our wide-area survey, our work shows that the PE galaxy SMF is fit very well by the Schechter functional form all the way up to very high stellar masses (log(\Mstars/\Msun)$>$11.5. The excellent Schechter-like PE galaxy SMF at \zs1.6, including at the high-mass end, leads to the immediate conclusion that the  mass quenching mechanism must have already been well established by \zs1.6, and earlier still if one assumes a non-zero timescale for galaxies to transition from the star-forming to the quiescent populations. Furthermore, given the excellent Schechter-like PE galaxy SMF, this mass quenching mechanism is likely due to a single physical process over a wide range of galaxy mass, including at ultra-high mass (log(\Mstars/\Msun)$>$11.5). 

Furthermore, again given the very good Schechter-like PE galaxy SMF, gravitational lensing, which is invoked to explain the non-Schechter luminosity functions sometimes observed at higher redshifts (e.g., \citealt{Hildebrandt2009, Ono2018}) is unlikely to be significant for massive \zs1.6 galaxies. 

As noted above, the mass quenching mechanism predicts a straightforward relation between the PE and SF Schechter functions: $M^*_{PE} = M^*_{SF}$ and $\alpha = \alpha_{PE}-\alpha_{SF} \sim 1$. Given the large dispersion in the measurements of $\alpha$ in the literature (see Table~\ref{tab:best-fit-params}), we are not in a position to test this prediction for the low-mass slopes.  However, the PE galaxy $M^*_{PE}$ values are fairly similar between the different surveys (even if often formally inconsistent given the stated uncertainties), so we can test whether $M^*_{PE} = M^*_{SF}$.
To this end, in Fig.~\ref{fig:MstarComparison} we compare measurements of  $M^*_{PE}$ and $M^*_{SF}$ at \zs1.5--2. We plot $M^{*}$ values from \cite{Ilbert2013b}, \cite{Muzzin2013smf}, and \cite{Tomczak2014} as well as our results (complemented with the $M^{*}_{SF}$ for star-forming galaxies from \citealt{Ilbert2013b}).  The diagonal line shows $M^*_{PE} = M^*_{SF}$.  It is clear that in most cases $M^*_{SF} > M^*_{PE}$ (the exception is the measurement of \citealt{Ilbert2013b}). The offset between $M^*_{SF}$ and $M^*_{PE}$ increases further by $\sim$0.1 dex (indicated by the blue arrow) when we correct for the fact that stellar masses of star-forming (but not quiescent) galaxies at high redshift are underestimated in most studies due to the effect of outshining. 
The blue arrow in Fig.~\ref{fig:MstarComparison} shows the outshining correction to star-forming galaxy masses inferred using  Eq.~6 of \cite{Sorba2018} for an estimated $SFR\sim10^{2.1}$ for a galaxy with \Mstars\ $=M^*_{SF} = 10^{11}M_{\odot}$ from \cite{Whitaker2014}. 

The offset between $M^*_{SF}$ and $M^*_{PE}$ seems significant, apparently in disagreement with the simple prediction of the mass quenching model of \cite{Peng2010}. The offset cannot be due to the effect of galaxy-galaxy merging as merging would drive $M^*_{PE}$ up, while keeping $M^*_{SF}$ constant (the latter because a merger involving SF galaxies would increase the mass of the merger product thus making it immediately more susceptible to quenching). Since this behaviour in $M^*_{SF}$ and $M^*_{PE}$ is opposite of what we observe, we conclude that major merging does not play a significant role at high redshift. 

Instead, the presence of the offset between $M^*_{SF}$ and $M^*_{PE}$ may suggest that we are seeing the efficiency of mass quenching evolve with time. This would mean that (i) the physical mechanism responsible for mass quenching is redshift-dependant (i.e., $\mu = \mu(z)$ in Eq.~23 of \citealt{Peng2010}), and (ii) that the mass quenching mechanism is not instantaneous but, instead, has a quenching time $\tau_{MQ}$ that is consistent with a quenching mechanism with long time-scales, which might be due to smooth feedback mechanism but also to viral-shock heating \citep{Moutard2016b}. Together, these two phenomena will cause the $M^*$ of both SF and PE populations to evolve with time, but with $M^*_{PE}$ lagging behind $M^*_{SF}$ by an offset related to the quenching timescale  $\tau_{MQ}$. 

\section{Conclusions}

In this paper, the first in a series on \zs1.6 massive quiescent galaxies, we introduced our large sample of \gzK-selected passive galaxies.  These data have allowed us to constrain the number counts and mass function of this population with unprecedented precision, spanning the mass range $10.25 \la $ log(\Mstars/\Msun)$ \la 11.75$. This mass range reaches extremely rare objects with ultra-high masses that were not previously included in significant numbers in smaller surveys. We find that our PE galaxy SMF can be well fit with a Schechter function over essentially this entire mass range.  The fact that the SMF has the Schechter form is consistent with the mass quenching scenario of \citep{Peng2010}; interpreting our result in the context of this scenario, we draw the following conclusions:

\begin{enumerate}
\item Given the SMF of PE galaxies is well represented by the Schechter function at \zs1.6, the mass quenching mechanism must have already been well established by this epoch.  It may well have been established earlier than that if the time to transition from the star-forming to the quenched population is significant. 

\item Given the adherence of the PE galaxy SMF to the Schechter function shape over a wide mass range, the mass quenching mechanism is likely a single physical process over a wide range of galaxy mass, including at ultra-high masses [log(\Mstars/\Msun)$>$11.5]. 

\item Gravitational lensing, which is invoked to explain departures from the Schechter shape at higher redshifts (e.g., \citealt{Hildebrandt2009,Ono2018}) is unlikely to be significant for massive \zs1.6 galaxies as it would distort our SMF away from its Schechter function shape. 

\item Comparing $M^*_{PE}$ and $M^*_{SF}$ measurements (from this study and from other surveys) we find $M^*_{SF} > M^*_{PE}$, in apparent conflict with the 
simple mass-quenching model.  This offset can, however, be explained if the quenching efficiency ($\mu$ in the notation of \citealt{Peng2010}) evolves with time.  Additionally, this explanation for the offset between $M^*_{PE}$ and $M^*_{SF}$ requires a slow quenching time-scale as the $M^*_{PE}$ lags behind $M^*_{SF}$. 

\item Galaxy-galaxy merging would distort the Schechter shape and would result in $M^*_{PE} > M^*_{SF}$, opposite to what is observed.  Merging is therefore not likely to be a strong mechanism for the growth of masses of ultra-massive galaxies at high redshift. 

\end{enumerate}

In addition to the results enumerated above, in this paper we have developed and described a very large sample of \zs1.6 quiescent galaxies. In forthcoming papers in this series we will study this massive quiescent population: we will study the clustering of  ultra-massive [log(\Mstars/\Msun) $>$ 11.5] members of the quiescent population (G.~Cheema, MNRAS, submitted) as well as their environments and growth through mergers (L.~Arcila-Osejo, in prep.). And we will identify and study protoclusters of quiescent galaxies found in our catalog (L.~Arcila-Osejo, in prep.). 

\section*{Acknowledgements}

We thank Ivana Damjanov, Laura Parker, and Rob Thacker for useful suggestions and the Natural Sciences and Engineering Research Council (NSERC) of Canada for financial support. 

This work is based on observations obtained with MegaPrime/MegaCam, a joint project of CFHT and CEA/DAPNIA, at the Canada-France-Hawaii Telescope (CFHT) which is operated by the National Research Council (NRC) of Canada, the Institut National des Science de l'Univers of the Centre National de la Recherche Scientifique (CNRS) of France, and the University of Hawaii. This work uses data products from TERAPIX and the Canadian Astronomy Data Centre. It makes use of the VIPERS-MLS database, operated at CeSAM/LAM, Marseille, France. This work is based in part on observations obtained with WIRCam, a joint project of CFHT, Taiwan, Korea, Canada and France.  The research was carried out using computing resources from ACEnet and Compute Canada.






\bibliographystyle{mnras}
\bibliography{Ref3.bib} 

\begin{thebibliography}{}
\makeatletter
\relax
\def\mn@urlcharsother{\let\do\@makeother \do\$\do\&\do\#\do\^\do\_\do\%\do\~}
\def\mn@doi{\begingroup\mn@urlcharsother \@ifnextchar [ {\mn@doi@}
  {\mn@doi@[]}}
\def\mn@doi@[#1]#2{\def\@tempa{#1}\ifx\@tempa\@empty \href
  {http://dx.doi.org/#2} {doi:#2}\else \href {http://dx.doi.org/#2} {#1}\fi
  \endgroup}
\def\mn@eprint#1#2{\mn@eprint@#1:#2::\@nil}
\def\mn@eprint@arXiv#1{\href {http://arxiv.org/abs/#1} {{\tt arXiv:#1}}}
\def\mn@eprint@dblp#1{\href {http://dblp.uni-trier.de/rec/bibtex/#1.xml}
  {dblp:#1}}
\def\mn@eprint@#1:#2:#3:#4\@nil{\def\@tempa {#1}\def\@tempb {#2}\def\@tempc
  {#3}\ifx \@tempc \@empty \let \@tempc \@tempb \let \@tempb \@tempa \fi \ifx
  \@tempb \@empty \def\@tempb {arXiv}\fi \@ifundefined
  {mn@eprint@\@tempb}{\@tempb:\@tempc}{\expandafter \expandafter \csname
  mn@eprint@\@tempb\endcsname \expandafter{\@tempc}}}

\bibitem[\protect\citeauthoryear{Arcila-Osejo \& Sawicki}{Arcila-Osejo \&
  Sawicki}{2013}]{Arcila-Osejo2013}
Arcila-Osejo L.,  Sawicki M.,  2013, \mn@doi [Monthly Notices of the Royal
  Astronomical Society] {10.1093/mnras/stu1356}, 435, 845

\bibitem[\protect\citeauthoryear{Behroozi \& Silk}{Behroozi \&
  Silk}{2018}]{Behroozi2018}
Behroozi P.,  Silk J.,  2018, \mn@doi [Monthly Notices of the Royal
  Astronomical Society] {10.1093/mnras/sty945}, 477, 5382

\bibitem[\protect\citeauthoryear{Belli, Newman  \& Ellis}{Belli
  et~al.}{2014a}]{Belli2014z1}
Belli S.,  Newman A.~B.,   Ellis R.~S.,  2014a, \mn@doi [The Astrophysical
  Journal] {10.1088/0004-637X/783/2/117}, 783, 117

\bibitem[\protect\citeauthoryear{Belli, Newman, Ellis  \& Konidaris}{Belli
  et~al.}{2014b}]{Belli2014z2}
Belli S.,  Newman A.~B.,  Ellis R.~S.,   Konidaris N.~P.,  2014b, \mn@doi [The
  Astrophysical Journal] {10.1088/2041-8205/788/2/L29}, 788, L29

\bibitem[\protect\citeauthoryear{Belli, Newman  \& Ellis}{Belli
  et~al.}{2016}]{Belli2016}
Belli S.,  Newman A.~B.,   Ellis R.~S.,  2016, \mn@doi [The Astrophysical
  Journal] {10.3847/1538-4357/834/1/18}, 834, 18

\bibitem[\protect\citeauthoryear{Belli, Newman  \& Ellis}{Belli
  et~al.}{2018}]{Belli2018}
Belli S.,  Newman A.~B.,   Ellis R.~S.,  2018, arXiv:1810.00008 [astro-ph]

\bibitem[\protect\citeauthoryear{Bertin \& Arnouts}{Bertin \&
  Arnouts}{1996}]{Bertin1996}
Bertin E.,  Arnouts S.,  1996, \mn@doi [Astronomy and Astrophysics Supplement
  Series] {10.1051/aas:1996164}, 117, 393

\bibitem[\protect\citeauthoryear{Bielby, Hudelot, McCracken, Ilbert, Daddi,
  F{\`{e}}vre  \& Kneib}{Bielby et~al.}{2012}]{Bielby2012}
Bielby R.,  Hudelot P.,  McCracken H.~J.,  Ilbert O.,  Daddi E.,  F{\`{e}}vre
  O.~L.,   Kneib J.,  2012, Astronomy \& Astrophysics, 23

\bibitem[\protect\citeauthoryear{{Bower}, {Benson}, {Malbon}, {Helly}, {Frenk},
  {Baugh}, {Cole}  \& {Lacey}}{{Bower} et~al.}{2006}]{Bower2006}
{Bower} R.~G.,  {Benson} A.~J.,  {Malbon} R.,  {Helly} J.~C.,  {Frenk} C.~S.,
  {Baugh} C.~M.,  {Cole} S.,   {Lacey} C.~G.,  2006, \mn@doi [Monthly Notices
  of the Royal Astronomical Society] {10.1111/j.1365-2966.2006.10519.x}, \href
  {http://adsabs.harvard.edu/abs/2006MNRAS.370..645B} {370, 645}

\bibitem[\protect\citeauthoryear{Bruzual \& Charlot}{Bruzual \&
  Charlot}{2003}]{Bruzual2003}
Bruzual G.,  Charlot S.,  2003, Monthly Notices of the Royal Astronomical
  Society, 344, 1000

\bibitem[\protect\citeauthoryear{Calzetti, Armus, Bohlin, Kinney, Koornneef  \&
  Storchi-Bergmann}{Calzetti et~al.}{2000}]{Calzetti2000a}
Calzetti D.,  Armus L.,  Bohlin R.~C.,  Kinney A.~L.,  Koornneef J.,
  Storchi-Bergmann T.,  2000, \mn@doi [The Astrophysical Journal]
  {10.1086/308692}, 533, 682

\bibitem[\protect\citeauthoryear{Capak et~al.,}{Capak et~al.}{2011}]{Capak2011}
Capak P.~L.,  et~al., 2011, \mn@doi [Nature] {10.1038/nature09681}, 470, 233

\bibitem[\protect\citeauthoryear{Chabrier}{Chabrier}{2003}]{Chabrier2003a}
Chabrier G.,  2003, \mn@doi [Publications of the Astronomical Society of the
  Pacific] {10.1086/376392}, 115, 763

\bibitem[\protect\citeauthoryear{Cimatti et~al.,}{Cimatti
  et~al.}{2004}]{Cimatti2004}
Cimatti A.,  et~al., 2004, ] {10.1038/nature02668}

\bibitem[\protect\citeauthoryear{Cimatti et~al.,}{Cimatti
  et~al.}{2008}]{Cimatti2008}
Cimatti A.,  et~al., 2008, \mn@doi [Astronomy and Astrophysics]
  {10.1051/0004-6361:20078739}, 482, 21–42

\bibitem[\protect\citeauthoryear{Cucciati et~al.,}{Cucciati
  et~al.}{2014}]{Cucciati2014}
Cucciati O.,  et~al., 2014, \mn@doi [Astronomy & Astrophysics]
  {10.1051/0004-6361/201423811}, 570, A16

\bibitem[\protect\citeauthoryear{Daddi, Cimatti, Renzini, Fontana, Mignoli,
  Pozzetti, Tozzi  \& Zamorani}{Daddi et~al.}{2004}]{Daddi2004b}
Daddi E.,  Cimatti A.,  Renzini A.,  Fontana A.,  Mignoli M.,  Pozzetti L.,
  Tozzi P.,   Zamorani G.,  2004, \mn@doi [The Astrophysical Journal]
  {10.1086/425569}, 617, 746

\bibitem[\protect\citeauthoryear{Daddi et~al.,}{Daddi et~al.}{2005}]{Daddi2005}
Daddi E.,  et~al., 2005, \mn@doi [The Astrophysical Journal] {10.1086/430104},
  626, 680

\bibitem[\protect\citeauthoryear{{Daddi} et~al.,}{{Daddi}
  et~al.}{2007}]{Daddi2007}
{Daddi} E.,  et~al., 2007, \mn@doi [The Astrophysical Journal]
  {10.1086/521818}, \href {http://adsabs.harvard.edu/abs/2007ApJ...670..156D}
  {670, 156}

\bibitem[\protect\citeauthoryear{Davidzon et~al.,}{Davidzon
  et~al.}{2017}]{Davidzon2017}
Davidzon I.,  et~al., 2017, \mn@doi [Astronomy and Astrophysics]
  {10.1051/0004-6361/201730419}, 605, A70

\bibitem[\protect\citeauthoryear{De~Lucia \& Blaizot}{De~Lucia \&
  Blaizot}{2007}]{DeLucia2007}
De~Lucia G.,  Blaizot J.,  2007, \mn@doi [Monthly Notices of the Royal
  Astronomical Society] {10.1111/j.1365-2966.2006.11287.x}, 375, 2

\bibitem[\protect\citeauthoryear{Dunlop, Peacock, Spinrad, Dey, Jimenez, Stern
  \& Windhorst}{Dunlop et~al.}{1996}]{Dunlop1996}
Dunlop J.,  Peacock J.,  Spinrad H.,  Dey A.,  Jimenez R.,  Stern D.,
  Windhorst R.,  1996, \mn@doi [Nature] {10.1038/381581a0}, 381, 581–584

\bibitem[\protect\citeauthoryear{{Eddington}}{{Eddington}}{1913}]{Eddington1913}
{Eddington} A.~S.,  1913, \mn@doi [\mnras] {10.1093/mnras/73.5.359}, \href
  {http://adsabs.harvard.edu/abs/1913MNRAS..73..359E} {73, 359}

\bibitem[\protect\citeauthoryear{Elbaz et~al.,}{Elbaz et~al.}{2007}]{Elbaz2007}
Elbaz D.,  et~al., 2007, \mn@doi [Astronomy & Astrophysics]
  {10.1051/0004-6361:20077525}, 468, 33

\bibitem[\protect\citeauthoryear{Elbaz et~al.,}{Elbaz et~al.}{2018}]{Elbaz2018}
Elbaz D.,  et~al., 2018, \mn@doi [Astronomy & Astrophysics]
  {10.1051/0004-6361/201732370}, 616, A110

\bibitem[\protect\citeauthoryear{Erben et~al.,}{Erben et~al.}{2012}]{Erben2012}
Erben T.,  et~al., 2012, Monthly Notices of the Royal Astronomical Society, 14,
  1

\bibitem[\protect\citeauthoryear{Fang, Ma, Chen  \& Kong}{Fang
  et~al.}{2015}]{Fang2015}
Fang G.-W.,  Ma Z.-Y.,  Chen Y.,   Kong X.,  2015, Research in Astronomy and
  Astrophysics, 15, 811

\bibitem[\protect\citeauthoryear{Furusawa, Sekiguchi, Takata, Furusawa,
  Shimasaku, Simpson  \& Akiyama}{Furusawa et~al.}{2011}]{Furusawa2011}
Furusawa J.,  Sekiguchi K.,  Takata T.,  Furusawa H.,  Shimasaku K.,  Simpson
  C.,   Akiyama M.,  2011, \mn@doi [The Astrophysical Journal]
  {10.1088/0004-637X/727/2/111}, 727, 111

\bibitem[\protect\citeauthoryear{Glazebrook et~al.,}{Glazebrook
  et~al.}{2004}]{Glazebrook2004}
Glazebrook K.,  et~al., 2004, \mn@doi [Nature] {10.1038/nature02667}, 430, 181

\bibitem[\protect\citeauthoryear{{Glazebrook} et~al.,}{{Glazebrook}
  et~al.}{2017}]{Glazebrook2017}
{Glazebrook} K.,  et~al., 2017, \mn@doi [\nat] {10.1038/nature21680}, \href
  {http://adsabs.harvard.edu/abs/2017Natur.544...71G} {544, 71}

\bibitem[\protect\citeauthoryear{Gobat et~al.,}{Gobat et~al.}{2012}]{Gobat2012}
Gobat R.,  et~al., 2012, \mn@doi [The Astrophysical Journal]
  {10.1088/2041-8205/759/2/L44}, 759, L44

\bibitem[\protect\citeauthoryear{Gobat et~al.,}{Gobat et~al.}{2013}]{Gobat2013}
Gobat R.,  et~al., 2013, \mn@doi [The Astrophysical Journal]
  {10.1088/0004-637X/776/1/9}, 776, 9

\bibitem[\protect\citeauthoryear{Goranova, Hudelot, McCracken  \&
  Mellier}{Goranova et~al.}{2010}]{Cuillandre2010}
Goranova Y.,  Hudelot P.,  McCracken H.~J.,   Mellier Y.,  2010, The CFHTLS
  T0006 release

\bibitem[\protect\citeauthoryear{{Goulding} et~al.,}{{Goulding}
  et~al.}{2012}]{Goulding2012}
{Goulding} A.~D.,  et~al., 2012, \mn@doi [The Astrophysical Journal,
  Supplement] {10.1088/0067-0049/202/1/6}, \href
  {http://adsabs.harvard.edu/abs/2012ApJS..202....6G} {202, 6}

\bibitem[\protect\citeauthoryear{{Hartley} et~al.,}{{Hartley}
  et~al.}{2008}]{Hartley2008}
{Hartley} W.~G.,  et~al., 2008, \mn@doi [\mnras]
  {10.1111/j.1365-2966.2008.13956.x}, \href
  {http://adsabs.harvard.edu/abs/2008MNRAS.391.1301H} {391, 1301}

\bibitem[\protect\citeauthoryear{Hayashi et~al.,}{Hayashi
  et~al.}{2009}]{Hayashi2009}
Hayashi M.,  et~al., 2009, \mn@doi [The Astrophysical Journal]
  {10.1088/0004-637X/691/1/140}, 691, 140

\bibitem[\protect\citeauthoryear{{Hildebrandt}, {van Waerbeke}  \&
  {Erben}}{{Hildebrandt} et~al.}{2009}]{Hildebrandt2009}
{Hildebrandt} H.,  {van Waerbeke} L.,   {Erben} T.,  2009, \mn@doi [Astronomy
  and Astrophysics] {10.1051/0004-6361/200912655}, \href
  {http://adsabs.harvard.edu/abs/2009A%26A...507..683H} {507, 683}

\bibitem[\protect\citeauthoryear{Hudelot, Goranova, Mellier, McCracken,
  Magnard, Monnerville, S{\'{e}}mah  \& Schultheis}{Hudelot
  et~al.}{2012}]{Hudelot2012a}
Hudelot P.,  Goranova Y.,  Mellier Y.,  McCracken H.~J.,  Magnard F.,
  Monnerville M.,  S{\'{e}}mah G.,   Schultheis M.,  2012, \mn@doi [Proceedings
  of the SPIE] {10.1117/12.925584}, 8448

\bibitem[\protect\citeauthoryear{Ilbert et~al.,}{Ilbert
  et~al.}{2013}]{Ilbert2013b}
Ilbert O.,  et~al., 2013, \mn@doi [Astronomy {\&} Astrophysics]
  {10.1051/0004-6361/201321100}, 556, A55

\bibitem[\protect\citeauthoryear{{Ishikawa}, {Kashikawa}, {Toshikawa}  \&
  {Onoue}}{{Ishikawa} et~al.}{2015}]{Ishikawa2015}
{Ishikawa} S.,  {Kashikawa} N.,  {Toshikawa} J.,   {Onoue} M.,  2015, \mn@doi
  [\mnras] {10.1093/mnras/stv1927}, \href
  {http://adsabs.harvard.edu/abs/2015MNRAS.454..205I} {454, 205}

\bibitem[\protect\citeauthoryear{{Ishikawa}, {Kashikawa}, {Hamana}, {Toshikawa}
   \& {Onoue}}{{Ishikawa} et~al.}{2016}]{Ishikawa2016}
{Ishikawa} S.,  {Kashikawa} N.,  {Hamana} T.,  {Toshikawa} J.,   {Onoue} M.,
  2016, \mn@doi [\mnras] {10.1093/mnras/stw271}, \href
  {http://adsabs.harvard.edu/abs/2016MNRAS.458..747I} {458, 747}

\bibitem[\protect\citeauthoryear{{Kado-Fong} et~al.,}{{Kado-Fong}
  et~al.}{2017}]{Kado-Fong2017}
{Kado-Fong} E.,  et~al., 2017, \mn@doi [\apj] {10.3847/1538-4357/aa6037}, \href
  {http://adsabs.harvard.edu/abs/2017ApJ...838...57K} {838, 57}

\bibitem[\protect\citeauthoryear{Kitzbichler \& White}{Kitzbichler \&
  White}{2007}]{Kitzbichler2007}
Kitzbichler M.~G.,  White S. D.~M.,  2007, \mn@doi [Monthly Notices of the
  Royal Astronomical Society] {10.1111/j.1365-2966.2007.11458.x}, 376, 2

\bibitem[\protect\citeauthoryear{Kurczynski et~al.,}{Kurczynski
  et~al.}{2012}]{Kurczynski2012}
Kurczynski P.,  et~al., 2012, \mn@doi [The Astrophysical Journal]
  {10.1088/0004-637X/750/2/117}, 750, 117

\bibitem[\protect\citeauthoryear{{Lee} et~al.,}{{Lee} et~al.}{2013}]{Lee2013}
{Lee} B.,  et~al., 2013, \mn@doi [\apj] {10.1088/0004-637X/774/1/47}, \href
  {http://adsabs.harvard.edu/abs/2013ApJ...774...47L} {774, 47}

\bibitem[\protect\citeauthoryear{Lidman et~al.,}{Lidman
  et~al.}{2012}]{Lidman2012}
Lidman C.,  et~al., 2012, \mn@doi [Monthly Notices of the Royal Astronomical
  Society] {10.1111/j.1365-2966.2012.21984.x}, 427, 550

\bibitem[\protect\citeauthoryear{{Lilly}, {Tresse}, {Hammer}, {Crampton}  \&
  {Le Fevre}}{{Lilly} et~al.}{1995}]{Lilly1995}
{Lilly} S.~J.,  {Tresse} L.,  {Hammer} F.,  {Crampton} D.,   {Le Fevre} O.,
  1995, \mn@doi [\apj] {10.1086/176560}, \href
  {http://adsabs.harvard.edu/abs/1995ApJ...455..108L} {455, 108}

\bibitem[\protect\citeauthoryear{Madau \& Dickinson}{Madau \&
  Dickinson}{2014}]{Madau2014}
Madau P.,  Dickinson M.,  2014, \mn@doi [Annual Review of Astronomy and
  Astrophysics] {10.1146/annurev-astro-081811-125615}, 52, 415

\bibitem[\protect\citeauthoryear{Mancini et~al.,}{Mancini
  et~al.}{2010}]{Mancini2010a}
Mancini C.,  et~al., 2010, \mn@doi [Monthly Notices of the Royal Astronomical
  Society] {10.1111/j.1365-2966.2009.15728.x}, 401, 933

\bibitem[\protect\citeauthoryear{{Marchesini}, {van Dokkum}, {F{\"o}rster
  Schreiber}, {Franx}, {Labb{\'e}}  \& {Wuyts}}{{Marchesini}
  et~al.}{2009}]{Marchesini2009}
{Marchesini} D.,  {van Dokkum} P.~G.,  {F{\"o}rster Schreiber} N.~M.,  {Franx}
  M.,  {Labb{\'e}} I.,   {Wuyts} S.,  2009, \mn@doi [The Astrophysical Journal]
  {10.1088/0004-637X/701/2/1765}, \href
  {http://adsabs.harvard.edu/abs/2009ApJ...701.1765M} {701, 1765}

\bibitem[\protect\citeauthoryear{{McCarthy}, {Babul}, {Bower}  \&
  {Balogh}}{{McCarthy} et~al.}{2008}]{McCarthy2008}
{McCarthy} I.~G.,  {Babul} A.,  {Bower} R.~G.,   {Balogh} M.~L.,  2008, \mn@doi
  [Montly Notices of the Royal Astronomical Society]
  {10.1111/j.1365-2966.2008.13141.x}, \href
  {http://adsabs.harvard.edu/abs/2008MNRAS.386.1309M} {386, 1309}

\bibitem[\protect\citeauthoryear{McCracken et~al.,}{McCracken
  et~al.}{2010}]{McCracken2010}
McCracken H.~J.,  et~al., 2010, \mn@doi [The Astrophysical Journal]
  {10.1088/0004-637X/708/1/202}, 708, 202

\bibitem[\protect\citeauthoryear{{McCracken} et~al.,}{{McCracken}
  et~al.}{2012}]{McCracken2012}
{McCracken} H.~J.,  et~al., 2012, \mn@doi [Astronomy and Astrophysics]
  {10.1051/0004-6361/201219507}, \href
  {http://adsabs.harvard.edu/abs/2012A%26A...544A.156M} {544, A156}

\bibitem[\protect\citeauthoryear{Merson et~al.,}{Merson
  et~al.}{2013}]{Merson2013}
Merson A.~I.,  et~al., 2013, \mn@doi [Monthly Notices of the Royal Astronomical
  Society] {10.1093/mnras/sts355}, 429, 556

\bibitem[\protect\citeauthoryear{Micha\l{}owski, Dunlop, Cirasuolo, Hjorth,
  Hayward  \& Watson}{Micha\l{}owski et~al.}{2012}]{Michalowski2012}
Micha\l{}owski M.~J.,  Dunlop J.~S.,  Cirasuolo M.,  Hjorth J.,  Hayward C.~C.,
    Watson D.,  2012, \mn@doi [Astronomy & Astrophysics]
  {10.1051/0004-6361/201016308}, 541, A85

\bibitem[\protect\citeauthoryear{Micha\l{}owski, Hayward, Dunlop, Bruce,
  Cirasuolo, Cullen  \& Hernquist}{Micha\l{}owski
  et~al.}{2014}]{Michalowski2014}
Micha\l{}owski M.~J.,  Hayward C.~C.,  Dunlop J.~S.,  Bruce V.~A.,  Cirasuolo
  M.,  Cullen F.,   Hernquist L.,  2014, \mn@doi [Astronomy & Astrophysics]
  {10.1051/0004-6361/201424174}, 571, A75

\bibitem[\protect\citeauthoryear{{Moutard} et~al.,}{{Moutard}
  et~al.}{2016a}]{Moutard2016a}
{Moutard} T.,  et~al., 2016a, \mn@doi [Astronomy \& Astrophysics]
  {10.1051/0004-6361/201527945}, \href
  {http://adsabs.harvard.edu/abs/2016A%26A...590A.102M} {590, A102}

\bibitem[\protect\citeauthoryear{{Moutard} et~al.,}{{Moutard}
  et~al.}{2016b}]{Moutard2016b}
{Moutard} T.,  et~al., 2016b, \mn@doi [Astronomy \& Astrophysics]
  {10.1051/0004-6361/201527294}, \href
  {http://adsabs.harvard.edu/abs/2016A%26A...590A.103M} {590, A103}

\bibitem[\protect\citeauthoryear{Muzzin et~al.,}{Muzzin
  et~al.}{2013a}]{Muzzin2013cat}
Muzzin A.,  et~al., 2013a, \mn@doi [The Astrophysical Journal Supplement
  Series] {10.1088/0067-0049/206/1/8}, 206, 8

\bibitem[\protect\citeauthoryear{Muzzin et~al.,}{Muzzin
  et~al.}{2013b}]{Muzzin2013smf}
Muzzin A.,  et~al., 2013b, \mn@doi [The Astrophysical Journal]
  {10.1088/0004-637X/777/1/18}, 777, 18

\bibitem[\protect\citeauthoryear{Newman, Ellis, Andreon, Treu, Raichoor  \&
  Trinchieri}{Newman et~al.}{2014}]{Newman2014}
Newman A.~B.,  Ellis R.~S.,  Andreon S.,  Treu T.,  Raichoor A.,   Trinchieri
  G.,  2014, \mn@doi [The Astrophysical Journal] {10.1088/0004-637X/788/1/51},
  788, 51

\bibitem[\protect\citeauthoryear{{Noeske} et~al.,}{{Noeske}
  et~al.}{2007}]{Noeske2007}
{Noeske} K.~G.,  et~al., 2007, \mn@doi [The Astrophysical Journal, Letters]
  {10.1086/517926}, \href {http://adsabs.harvard.edu/abs/2007ApJ...660L..43N}
  {660, L43}

\bibitem[\protect\citeauthoryear{{Oke}}{{Oke}}{1974}]{Oke1974}
{Oke} J.~B.,  1974, \mn@doi [\apjs] {10.1086/190287}, \href
  {http://adsabs.harvard.edu/abs/1974ApJS...27...21O} {27, 21}

\bibitem[\protect\citeauthoryear{{Ono} et~al.,}{{Ono} et~al.}{2018}]{Ono2018}
{Ono} Y.,  et~al., 2018, \mn@doi [Publications of the Astronomical Society of
  Japan] {10.1093/pasj/psx103}, \href
  {http://adsabs.harvard.edu/abs/2018PASJ...70S..10O} {70, S10}

\bibitem[\protect\citeauthoryear{Onodera, Arimoto, Daddi, Renzini, Kong,
  Cimatti, Broadhurst  \& Alexander}{Onodera et~al.}{2010}]{Onodera2010}
Onodera M.,  Arimoto N.,  Daddi E.,  Renzini A.,  Kong X.,  Cimatti A.,
  Broadhurst T.,   Alexander D.~M.,  2010, \mn@doi [The Astrophysical Journal]
  {10.1088/0004-637X/715/1/385}, 715, 385

\bibitem[\protect\citeauthoryear{Onodera et~al.,}{Onodera
  et~al.}{2012}]{Onodera2012}
Onodera M.,  et~al., 2012, \mn@doi [The Astrophysical Journal]
  {10.1088/0004-637X/755/1/26}, 755, 26

\bibitem[\protect\citeauthoryear{Oteo et~al.,}{Oteo et~al.}{2018}]{Oteo2018}
Oteo I.,  et~al., 2018, \mn@doi [The Astrophysical Journal]
  {10.3847/1538-4357/aaa1f1}, 856, 72

\bibitem[\protect\citeauthoryear{{Pacifici} et~al.,}{{Pacifici}
  et~al.}{2016}]{Pacifici2016}
{Pacifici} C.,  et~al., 2016, \mn@doi [The Astrophysical Journal]
  {10.3847/0004-637X/832/1/79}, \href
  {http://adsabs.harvard.edu/abs/2016ApJ...832...79P} {832, 79}

\bibitem[\protect\citeauthoryear{Peng et~al.,}{Peng et~al.}{2010}]{Peng2010}
Peng Y.-J.,  et~al., 2010, \mn@doi [The Astrophysical Journal]
  {10.1088/0004-637X/721/1/193}, 721, 193

\bibitem[\protect\citeauthoryear{{Peng}, {Lilly}, {Renzini}  \&
  {Carollo}}{{Peng} et~al.}{2012}]{Peng2012}
{Peng} Y.-j.,  {Lilly} S.~J.,  {Renzini} A.,   {Carollo} M.,  2012, \mn@doi
  [The Astrophysical Journal] {10.1088/0004-637X/757/1/4}, \href
  {http://adsabs.harvard.edu/abs/2012ApJ...757....4P} {757, 4}

\bibitem[\protect\citeauthoryear{Rangel, Nandra, Laird  \& Orange}{Rangel
  et~al.}{2013}]{Rangel2013}
Rangel C.,  Nandra K.,  Laird E.~S.,   Orange P.,  2013, \mn@doi [Monthly
  Notices of the Royal Astronomical Society] {10.1093/mnras/sts256}, 428, 3089

\bibitem[\protect\citeauthoryear{Ravindranath, Daddi, Giavalisco, Ferguson  \&
  Dickinson}{Ravindranath et~al.}{2007}]{Ravindranath2007}
Ravindranath S.,  Daddi E.,  Giavalisco M.,  Ferguson H.~C.,   Dickinson M.~E.,
   2007, \mn@doi [Proceedings of the International Astronomical Union]
  {10.1017/S1743921308018243}, 3, 407

\bibitem[\protect\citeauthoryear{{Sato}, {Sawicki}  \& {Arcila-Osejo}}{{Sato}
  et~al.}{2014}]{Sato2014}
{Sato} T.,  {Sawicki} M.,   {Arcila-Osejo} L.,  2014, \mn@doi [Monthly Notices
  of the Royal Astronomical Society] {10.1093/mnras/stu1356}, \href
  {http://adsabs.harvard.edu/abs/2014MNRAS.443.2661S} {443, 2661}

\bibitem[\protect\citeauthoryear{Sawicki}{Sawicki}{2012}]{Sawicki2012hdf}
Sawicki M.,  2012, \mn@doi [Monthly Notices of the Royal Astronomical Society]
  {10.1111/j.1365-2966.2012.20452.x}, 421, 2187

\bibitem[\protect\citeauthoryear{{Sawicki} \& {Thompson}}{{Sawicki} \&
  {Thompson}}{2006}]{Sawicki2006}
{Sawicki} M.,  {Thompson} D.,  2006, \mn@doi [\apj] {10.1086/500999}, \href
  {http://adsabs.harvard.edu/abs/2006ApJ...642..653S} {642, 653}

\bibitem[\protect\citeauthoryear{Sawicki \& Yee}{Sawicki \&
  Yee}{1998}]{Sawicki1998}
Sawicki M.,  Yee H. K.~C.,  1998, \mn@doi [The Astronomical Journal]
  {10.1086/300291}, 115, 1329

\bibitem[\protect\citeauthoryear{{Sawicki}, {Lin}  \& {Yee}}{{Sawicki}
  et~al.}{1997}]{Sawicki1997}
{Sawicki} M.~J.,  {Lin} H.,   {Yee} H.~K.~C.,  1997, \mn@doi [\aj]
  {10.1086/118231}, \href {http://adsabs.harvard.edu/abs/1997AJ....113....1S}
  {113, 1}

\bibitem[\protect\citeauthoryear{{Sawicki} et~al.,}{{Sawicki}
  et~al.}{2007}]{Sawicki2007}
{Sawicki} M.,  et~al., 2007, in {Afonso} J.,  {Ferguson} H.~C.,  {Mobasher} B.,
    {Norris} R.,  eds,  Astronomical Society of the Pacific Conference Series
  Vol. 380, Deepest Astronomical Surveys. p.~433 (\mn@eprint {}
  {astro-ph/0612117})

\bibitem[\protect\citeauthoryear{{Schechter}}{{Schechter}}{1976}]{Schechter1976}
{Schechter} P.,  1976, \mn@doi [The Astrophysical Journal] {10.1086/154079},
  \href {http://adsabs.harvard.edu/abs/1976ApJ...203..297S} {203, 297}

\bibitem[\protect\citeauthoryear{{Schlafly} \& {Finkbeiner}}{{Schlafly} \&
  {Finkbeiner}}{2011}]{Schlafly2011}
{Schlafly} E.~F.,  {Finkbeiner} D.~P.,  2011, \mn@doi [The Astrophysical
  Journal] {10.1088/0004-637X/737/2/103}, \href
  {http://adsabs.harvard.edu/abs/2011ApJ...737..103S} {737, 103}

\bibitem[\protect\citeauthoryear{{Schlegel}, {Finkbeiner}  \&
  {Davis}}{{Schlegel} et~al.}{1998}]{Schlegel1998}
{Schlegel} D.~J.,  {Finkbeiner} D.~P.,   {Davis} M.,  1998, \mn@doi [The
  Astrophysical Journal] {10.1086/305772}, \href
  {http://adsabs.harvard.edu/abs/1998ApJ...500..525S} {500, 525}

\bibitem[\protect\citeauthoryear{Somerville, Gilmore, Primack  \&
  Dom{\'{i}}nguez}{Somerville et~al.}{2012}]{Somerville2012}
Somerville R.~S.,  Gilmore R.~C.,  Primack J.~R.,   Dom{\'{i}}nguez A.,  2012,
  \mn@doi [Monthly Notices of the Royal Astronomical Society]
  {10.1111/j.1365-2966.2012.20490.x}, 423, 1992

\bibitem[\protect\citeauthoryear{{Sommariva} et~al.,}{{Sommariva}
  et~al.}{2014}]{Sommariva2014}
{Sommariva} V.,  et~al., 2014, \mn@doi [Astronomy \& Astrophysics]
  {10.1051/0004-6361/201322301}, \href
  {http://adsabs.harvard.edu/abs/2014A%26A...571A..99S} {571, A99}

\bibitem[\protect\citeauthoryear{{Sorba} \& {Sawicki}}{{Sorba} \&
  {Sawicki}}{2018}]{Sorba2018}
{Sorba} R.,  {Sawicki} M.,  2018, \mn@doi [Montly Notices of the Royal
  Astronomical Society] {10.1093/mnras/sty186}, \href
  {http://adsabs.harvard.edu/abs/2018MNRAS.476.1532S} {476, 1532}

\bibitem[\protect\citeauthoryear{{Springel} et~al.,}{{Springel}
  et~al.}{2005}]{Springel2005}
{Springel} V.,  et~al., 2005, \mn@doi [Nature] {10.1038/nature03597}, \href
  {http://adsabs.harvard.edu/abs/2005Natur.435..629S} {435, 629}

\bibitem[\protect\citeauthoryear{{Steidel}, {Adelberger}, {Dickinson},
  {Giavalisco}, {Pettini}  \& {Kellogg}}{{Steidel} et~al.}{1998}]{Steidel1998}
{Steidel} C.~C.,  {Adelberger} K.~L.,  {Dickinson} M.,  {Giavalisco} M.,
  {Pettini} M.,   {Kellogg} M.,  1998, \mn@doi [The Astrophysical Journal]
  {10.1086/305073}, \href {http://adsabs.harvard.edu/abs/1998ApJ...492..428S}
  {492, 428}

\bibitem[\protect\citeauthoryear{{Steidel}, {Adelberger}, {Giavalisco},
  {Dickinson}  \& {Pettini}}{{Steidel} et~al.}{1999}]{Steidel1999}
{Steidel} C.~C.,  {Adelberger} K.~L.,  {Giavalisco} M.,  {Dickinson} M.,
  {Pettini} M.,  1999, \mn@doi [\apj] {10.1086/307363}, \href
  {http://adsabs.harvard.edu/abs/1999ApJ...519....1S} {519, 1}

\bibitem[\protect\citeauthoryear{Steidel, Adelberger, Shapley, Pettini,
  Dickinson  \& Giavalisco}{Steidel et~al.}{2000}]{Steidel2000}
Steidel C.~C.,  Adelberger K.~L.,  Shapley A.~E.,  Pettini M.,  Dickinson M.,
  Giavalisco M.,  2000, \mn@doi [The Astrophysical Journal] {10.1086/308568},
  532, 170

\bibitem[\protect\citeauthoryear{Stott et~al.,}{Stott et~al.}{2010}]{Stott2010}
Stott J.~P.,  et~al., 2010, \mn@doi [The Astrophysical Journal]
  {10.1088/0004-637X/718/1/23}, 718, 23

\bibitem[\protect\citeauthoryear{Tomczak et~al.,}{Tomczak
  et~al.}{2014}]{Tomczak2014}
Tomczak A.,  et~al., 2014, \mn@doi [The Astrophysical Journal]
  {10.1088/0004-637X/783/2/85}, 783, 85

\bibitem[\protect\citeauthoryear{Wellons et~al.,}{Wellons
  et~al.}{2015}]{Wellons2015}
Wellons S.,  et~al., 2015, \mn@doi [Monthly Notices of the Royal Astronomical
  Society] {10.1093/mnras/stv303}, 449, 361

\bibitem[\protect\citeauthoryear{{Whitaker}, {van Dokkum}, {Brammer}  \&
  {Franx}}{{Whitaker} et~al.}{2012}]{Whitaker2012}
{Whitaker} K.~E.,  {van Dokkum} P.~G.,  {Brammer} G.,   {Franx} M.,  2012,
  \mn@doi [The Astrophysical Journal, Letters] {10.1088/2041-8205/754/2/L29},
  \href {http://adsabs.harvard.edu/abs/2012ApJ...754L..29W} {754, L29}

\bibitem[\protect\citeauthoryear{{Whitaker} et~al.,}{{Whitaker}
  et~al.}{2014}]{Whitaker2014}
{Whitaker} K.~E.,  et~al., 2014, \mn@doi [The Astrophysical Journal]
  {10.1088/0004-637X/795/2/104}, \href
  {http://adsabs.harvard.edu/abs/2014ApJ...795..104W} {795, 104}

\bibitem[\protect\citeauthoryear{{Yuma}, {Ohta}, {Yabe}, {Kajisawa}  \&
  {Ichikawa}}{{Yuma} et~al.}{2011}]{Yuma2011}
{Yuma} S.,  {Ohta} K.,  {Yabe} K.,  {Kajisawa} M.,   {Ichikawa} T.,  2011,
  \mn@doi [\apj] {10.1088/0004-637X/736/2/92}, \href
  {http://adsabs.harvard.edu/abs/2011ApJ...736...92Y} {736, 92}

\bibitem[\protect\citeauthoryear{Yuma, Ohta  \& Yabe}{Yuma
  et~al.}{2012}]{Yuma2012}
Yuma S.,  Ohta K.,   Yabe K.,  2012, \mn@doi [The Astrophysical Journal]
  {10.1088/0004-637X/761/1/19}, 761, 19

\makeatother
\end{thebibliography}



%
%


\bsp	
\label{lastpage}
\end{document}